\documentclass[showpacs,preprint,amsmath,amssymb,floatfix]{revtex4}

\usepackage{graphicx} 
\usepackage{dcolumn}           
\usepackage{bm}                

\begin{document}
\title{Effect of chain stiffness on ion distributions around a polyelectrolyte
       in multivalent salt solutions}
\author{Yu-Fu Wei} 
\author{Pai-Yi Hsiao} 
\email[corresponding author; e-mail: ]{pyhsiao@ess.nthu.edu.tw}
\affiliation{Department of Engineering and System Science, 
National Tsing Hua University, Hsinchu, Taiwan 300, R.O.C.}
\date{\today} 

\begin{abstract}
Ion distributions in dilute polyelectrolyte solutions are studied by means of
Langevin dynamics simulations. We show that the distributions depend on the
conformation of a chain while the conformation is determined by the chain
stiffness and the salt concentration. We observe that the monovalent
counterions originally condensed on a chain can be replaced by the multivalent
ones dissociated from the added salt due to strong electrostatic interaction.
These newly condensed ions give an important impact on the chain structure. At
low and at high salt concentrations, the conformation of a semiflexible chain
is rodlike. The ion distributions show similarity to those for a rigid chain,
but difference to those for a flexible chain whose conformation is a coil. In
the mid-salt region, the flexible chain and the semiflexible chain collapse but
the collapsed chain structures are, respectively, disordered and ordered
structures.  The ion distributions hence show different profiles for these
three chain stiffness with the curves for the semiflexible chain lying between
those for the flexible and the rigid chains. The number of the condensed
multivalent counterions, as well as the effective chain charge, also shows
similar behavior, demonstrating a direct connection with the chain
morphology. Moreover, we find that the condensed multivalent counterions form
triplets with two adjacent monomers and are localized on the chain axis at
intermediate salt concentration when the chain stiffness is semiflexible or
rigid. The microscopic information obtained here provides valuable insight to
the phenomena of DNA condensation and is very useful for researchers to develop
new models. 
 
\end{abstract}

\maketitle
\section{Introduction}
Polyelectrolytes (PEs), such as DNA molecules and many synthetic polymers, are
macromolecules carrying a large number of charged groups which are ionizable in
water. The presence of long-ranged electrostatic interaction and the
confinement of charges on chains, together with the degree of freedom of chain
conformation, make this kind of system very complicated and unpredictable.  The
behavior of PEs is greatly influenced by ions in the solutions, owing to the
screening effect and ionic redistribution. For example, addition of salt in the
solutions, especially multivalent salt, can change drastically the properties
of PEs~\cite{gelbart00,grosberg02,levin02,wong06,iwaki07}. When salt is added,
the multivalent counterions dissociated from the salt can compete with the
monovalent ones in the solutions and condense onto PE chains, which eventually
leads the system to a series of phase transition. The condensation of
multivalent counterions is a process of both energetic and entropic favor
because a $Z$-valent counterion ($Z>1$) attracting to a PE chain is $Z$ times
stronger than a monovalent one and the release of $Z$ monovalent counterions
into the solution, due to the condensation of the $Z$-valent counterion, also
increases the system entropy~\cite{bloomfield96,bloomfield97,messina09}.  A
phase separation takes place when salt concentration is higher than a critical
value. The chains collapse and aggregate in the solution, resulting in a
precipitation from the solution.  When the salt concentration is further
increased and goes beyond a second critical value, the precipitated PEs
redissolve back into the solution. The system returns to a homogeneous
phase~\cite{delacruz95,pelta96,raspaud98}. The phase transition occurred at the first
critical salt concentration is a result of ionic screening and bridging of
multivalent counterions bound near PEs~\cite{oosawa71,rouzina96,levin02}.  The
transition occurred at the second one exists several
explanations~\cite{solis00,solis01,solis02,nguyen00,grosberg02}. One of them
attributes the resolubilization of PEs to the occurrence of charge inversion of
PEs~\cite{nguyen00}. It argues that the condensed multivalent counterions can
overcompensate the chain charge when salt concentration is high, and the
repulsive force between chain segments due to the charge of inverted sign at
that moment causes the segregation and resolubilization of chains.  Another
theory considers the Bjerrum association between ions~\cite{solis02}.  By using
a two-state model of chain, it predicts the possibility of either sign of the
effective chain charge under different conditions. These theories usually
over-simplifies the situations and are constructed based upon certain
preassumed chain conformation and ion distribution. Further verification is
thus needed. To understand the systems, it is important to study chain
conformation and ion distributions around a chain in solutions. We have
investigated the first issue, the effect of chain stiffness on PE chain
conformation, in our previous paper~\cite{wei07}.  In this work, we focus on
the second issue: the ion distribution. 

To investigate ion distributions directly from experiments is difficult.
Computer simulation, on the contrary, can do it easily because the positions of
atoms are explicitely given in the study. This issue can be investigated on
different scales~\cite{young97,feig98,ponomarev04,abascal01,
abascal06,allahyarov03,allahyarov04,liu02,liu03,ou05,sarraguca03,sarraguca06,
klos05a,hsiao06a,hsiao06b}. An all-atom model, which simulates molecules with
detailed chemical composition and structure, are usually applied in the study
of small systems for short simulation time, owing to high computational cost
related to today's computer power~\cite{young97,feig98,ponomarev04}.  Space and
time scales are both restricted.  A coarse-grained model, which reduces the
complexity of the molecular structure by grouping atoms, on the other hands,
can decrease significantly the computational cost. Researchers can hence
investigate larger systems evolving over a longer simulation time. Using
coarse-grained models, scientists have investigated ion distribution and
effective interaction and free energy of DNA molecule
systems~\cite{abascal01,abascal06,allahyarov03,allahyarov04}. The chains were
modeled by an infinitively long post through periodic boundary condition. This
model preserved the DNA helical structure and can be used to study local
structural transitions, such as the B-to-Z transition, in addition to the ion
distribution. However, it fails to characterize the global chain conformation,
as what happened in DNA condensation when polyvalent ions are added into the
solution. 

To study the collapse and the reexpansion of single PE chains, bead-spring
chain models are usually
employed~\cite{liu02,liu03,ou05,sarraguca03,sarraguca06,
klos05a,hsiao06a,hsiao06b}. This kind of simple models allows people to
investigate the behavior of PE systems in a large space and time scale. By
exploiting it, researchers have demonstrated that various conformational
transitions of PE chains take places under different conditions. The models
could be hence good candidates in the study of ion distribution related to the
chain conformations. Liu and Muthukumar showed that a chain can be collapsed in
a salt-free solution when Coulomb coupling strength is strong~\cite{liu02}.
Their study reveals the relation between counterion condensation and the
formation of local dipole moment. Hsiao and Luijten studied how PE chains are
collapsed in multivalent salt solutions~\cite{hsiao06c}. They found that the
chains are charge overcompensated at their very surface when the amount of the
multivalent counterions of salt is greater than the amount needed to neutralize
the bare chain charge~\cite{hsiao06c}. Counterion condensation on a PE chain
has also been studied by several groups~\cite{limbach01,liu02,sarraguca06}.
However, these studies aimed to understand the collapse transition of chain
happened in salt-free or low-salt conditions. The information concerning the
unfolding transition of chain provoked by high concentration of counterions was
missing. In this work, we plane to study the ion distribution around a PE chain
in solutions over a wide range of salt concentration, so wide that single
chains can subsequently undergo collapse and reexpansion transitions, using
a bead-spring chain model.  The focus will be the exploitation of the ion
distribution at different salt concentrations and the relation with the chain
conformation. 

It is known that the structure of collapsed PE chains depends sensitively on
the chain rigidity~\cite{arscott90,plum90,maurstad03}. Studies also showed that
the rigidity determines the transition of chain conformation to be continuous
or discrete~\cite{lifshitz78,post82,ghosh02,yoshikawa95,yoshikawa96,wei07}. It
is, therefore, very relevant to know how ion distributions are influenced by
the chain rigidity in the problem of PE condensation. Ou and Muthukumar
observed that the number of the condensed monovalent counterions on a chain
decreases in a salt-free solution when chain stiffness increases~\cite{ou05}.
Sarragu\c{c}a and Pais studied PE chains collapse by trivalent salt in
different circumstances~\cite{sarraguca06}. They found that the density of
trivalent ions around stiff chains is smaller than around flexible ones.
Moreover, flexible chains are compacted into smaller dimensions and the ion
correlation is strong due to small volume of the confinement. Nonetheless, the
situation for high salt concentrations where chain reexpansion occurs was not
discussed.
 
This paper is organized as follows. We describe our model and the simulation
setup in Section~\ref{sec_model_setup}.  The results and discussions are given
in Section~\ref{sec_results}. After a short revisit of the dependence of  
chain size and morphology on the salt concentrations for flexible, semiflexible
and rigid chains in Section~\ref{subsec_chain_size}, we investigate various ion
distributions in PE solutions. The topics include  the integrated ion
distribution in the radial direction of a chain
(Section~\ref{subsec_ion_dist_in_chain_radial_direction}), the charge
distribution (Section~\ref{subsec_charge_dist_in_chain_radial_direction}), the
ion condensation and the effective chain charge
(Section~\ref{subsec_ion_condensation_&_effective_charge}), the condensed ion
distribution along a chain backbone
(Section~\ref{subsec_dist_cond_ion_along_chain}), and the radial distribution
functions (Section~\ref{subsec_rdf}).  By investigating these topics, we can
make connection between ion distributions and chain conformations and
understand the role of chain stiffness in different salt regions.  Moreover,
the results obtained from the rigid-chain case provide a direct verification of
the Manning condensation theory by simulations.  We give our conclusions in
Section~\ref{subsec_conclusions}.

\section{Model and setup}
\label{sec_model_setup}
Our system consists of a PE chain and tetravalent salt molecules, placed in a
cubic box with periodic boundary condition. The chain dissociates into a
polyion and many counterions.  The polyion is modeled by a bead-spring chain
and comprises 48 beads. Each bead carries a negative unit charge $-e$.  The
counterions are modeled by charged spheres, carrying a $+e$ charge each.  The
tetravalent salt dissociates into tetravalent cation (counterions) and
monovalent anions (coions). These ions are also modeled by charged spheres.
Solvent molecules are not explicitly incorporated in the study. Nonetheless, we
consider the entire solvent as a dielectric medium with dielectric constant
equal to $\epsilon_r$. 

We consider two kinds of non-bonded interactions: the excluded volume
interaction and the Coulomb interaction, applied to all the beads on the chain
and spheres in the system, and two kinds of bonded interactions: the
connectivity and the chain rigidity, applied solely to the chain.  The excluded
volume is modeled by a purely repulsive Lennard-Jones (LJ) potential
\begin{equation}
U_{LJ}(r)=\left\{\begin{array}{ll} 4\varepsilon_{LJ} \left[
\left(\sigma/r\right)^{12}- \left(\sigma/r\right)^6 \right] +\varepsilon_{LJ} &
\mbox{for } r \leq 2^{1/6} \sigma\\ 0 & \mbox{for } r > 2^{1/6} \sigma
\end{array}\right.
\end{equation}
where $r$ is the distance, $\varepsilon_{LJ}$ is the interaction strength and
$\sigma$ represents the diameter of a particle.  We assume identical
$\varepsilon_{LJ}$ and $\sigma$ for the ion spheres and monomer beads.  The
Coulomb interaction between two charged ions or beads is given by
\begin{equation}
U_{coul}(r)=\frac{Z_i Z_j \lambda_B k_B T}{r} \label{eqcoulomb2}
\end{equation}
where $Z_i$ and $Z_j$ are charge valences, $k_{B}$ is the Boltzmann constant,
$T$ is the temperature, and $\lambda_B=e^2/(4\pi\epsilon_0\epsilon_r k_B T)$ is
the Bjerrum length ($\epsilon_0$ is the vacuum permittivity and $\epsilon_r$
the solvent dielectric constant). The connectivity of two consecutive beads on
the chain is described by a finitely extensible nonlinear elastic potential
\begin{equation}
U_{sp}(b)= -\frac{1}{2} k_{b} b_{max}^2 \ln
\left(1-\frac{b^2}{b_{max}^2}\right)
\end{equation}
where $b$ is the length of a bond, $b_{max}$ is the maximum bond extention, and
$k_{b}$ is the spring constant. The chain rigidity is represented by a harmonic
angle potential,
\begin{equation}
U_{angle}(\theta)= k_{a}( \theta - \pi )^2
\end{equation}
where $\theta$ is the bond angle between two consecutive bonds and $k_{a}$ is
the force constant.

In addition to the above interactions, two more forces are exerted on each
particle. They are (1) a dissipative force, which takes into account the
friction force due to the particle moving through the solvent, and (2) a
stochastic force, which models the thermal collisions by solvent molecules.
The equation of motion, also known as the Langevin equation, reads as
\begin{equation}
m\ddot{\vec{r}}_i = -\frac{\partial U}{\partial \vec{r}_i} -m\gamma
\dot{\vec{r}}_i +\vec{{\eta}}_i(t) \label{eqlangevin}
\end{equation}
where $m$ is the mass of particle, $\gamma$ is the friction coefficient, and
$\vec{\eta}_i$ is the stochastic force.  We have assumed identical $m$ and
$\gamma$ for all the particles.  The system reaches an equilibrium state with
the temperature given by the fluctuation-dissipation theorem:
\begin{equation}
\left< \vec{\eta}_i(t) \cdot \vec{\eta}_j(t') \right> = 6 k_B T
m\gamma\delta_{ij} \delta(t-t') \label{eqflucdiss}
\end{equation}
where $\delta_{ij}$ and $\delta(t-t')$ are the Kronecker and the Dirac delta
functions, respectively.

We set $\varepsilon_{LJ}=k_BT/1.2$, $k_{b}=5.8333 k_B T/\sigma^2$,
$b_{max}=2\sigma$ and $\lambda_B=3\sigma$.  $\gamma$ is set to $15 \tau^{-1}$
(where $\tau=\sigma\sqrt{m/(k_B T)}$) to mimic aqueous solutions. These setups
give a linear charge density of chain corresponding to a value typically for
many polyelectrolyte systems, such as sodium polystyrene sulfonate. We varied
$k_{a}$ over a large range from 0 to $100 k_B T/rad^2$ to study the effect of
chain stiffness. The values cover three typical chain stiffness: flexible
chains, semiflexible chains, and rigid chains (see explanation in
Section~\ref{subsec_chain_size}). The (4:1)-salt concentration $C_s$ is varied
from $0$ to $0.01024 \sigma^{-3}$ whereas the monomer concentration $C_m$ is
fixed at $3.2 \times 10^{-4} \sigma^{-3}$.  The size of the simulation box is
$53.13 \sigma$, which is large enough to avoid chain overlapping under periodic
boundary condition. Therefore, the studied case is a dilute PE solution. It 
has been shown in experiments~\cite{pelta96} and simulations~\cite{hsiao06b} 
that PEs show similar condensation and decondensation behavior in the dilute
region.  The largest system in our study contains totally 7776
charged particles. We list all the values of $C_s$ investigated and the
corresponding numbers of tetravalent counterions and monovalent coions in
Table~\ref{tab:ion_no} for reader's reference.
\begin{table}[ht]
\centering
\begin{tabular}{c c c}
\hline\hline
$C_s\ (\times 10^{-5})$ & $\#(+4)$-ions & $\#(-1)$-ions \\
\hline
0.667 & 1 & 4 \\
1.33  & 2 & 8 \\
2.67  & 4 & 16 \\
4.00  & 6 & 24 \\
5.33  & 8 & 32 \\
8.00  & 12 & 48 \\
16.0  & 24 & 96 \\
32.0  & 48 & 192 \\
64.0  & 96 & 384 \\
128   & 192 & 768 \\
256   & 384 & 1536 \\
512   & 768 & 3072 \\
1024  & 1536 & 6144 \\
\hline
\end{tabular}
\caption{Values of $C_s$ and the corresponding numbers of tetravalent
counterions and monovalent coions in the simulation box}
\label{tab:ion_no}
\end{table}

We perform Langevin dynamics simulations with the integrating time step equal
to $0.005 \tau$. The Coulomb interaction is calculated by the technique of
Ewald sum. An equilibration phase takes about $10^7$ time steps of run,
followed by a production phase cumulating data for $10^7$ to $3\times10^8$ time
steps.  In order to sample sufficiently the data, several independent runs are
performed at each salt concentrations, starting with different initial
configurations. To simplify the notation, in the rest of the text,
length, energy, and time will be measured in unit of $\sigma$, $k_B T$, and
$\tau$, respectively. Therefore, concentration will be described in unit of
$\sigma^{-3}$ and $k_a$ in unit of $k_B T/rad^2$, and so on.

\section{Result and Discussion}
\label{sec_results}
\subsection{Chain size, morphology, and bond length}
\label{subsec_chain_size} In this section, we shortly revisit some of the major
findings obtained in our previous work~\cite{wei07}, how the chain size and
morphology depend on the chain stiffness in multivalent salt solutions, to give
readers a thorough picture of the system. The detailed discussions can be found
in the paper~\cite{wei07}. Starting from the next section, we will study ion
distributions and show that they depend strongly on the size and the morphology
of PE chains in solutions. 

\begin{figure} \centering
\includegraphics[height=0.7\textwidth,angle=270]{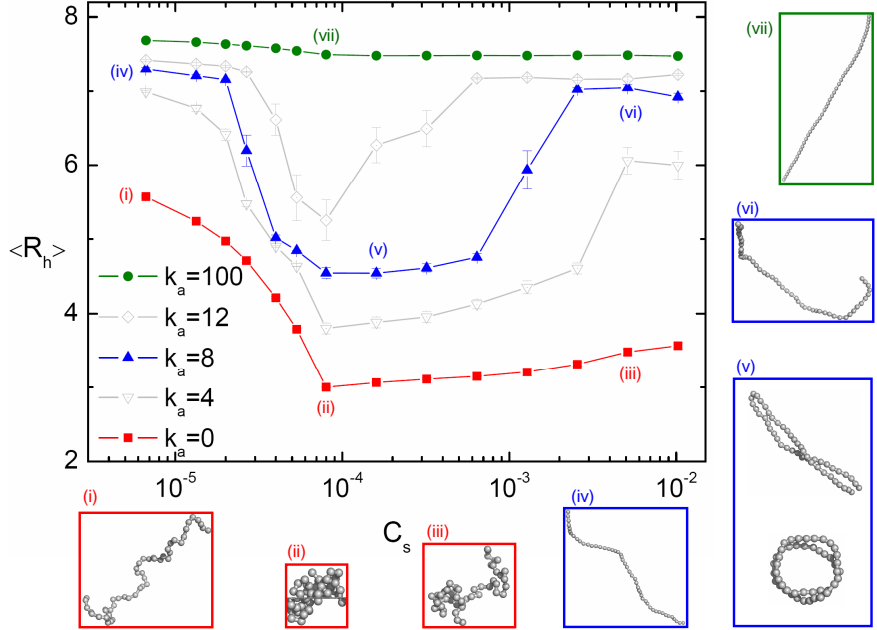} \caption{Mean
hydrodynamic radius $\langle R_h \rangle$ as a function of tetravalent salt
concentration $C_s$. Snapshots of some typical chain structures at the points
(i) to (vii) are shown by the side of the figure.} \label{f:RhvCs} \end{figure}

Instead by using the radius of gyration, we characterize here the chain size by the
hydrodynamic radius, $R_h$,  which is the Stokes radius of a polymer moving through 
a solution~\cite{strobl97}. $R_h$ is calculated by the equation $R_h^{-1}=
(\Sigma_{i=1}^{N}\Sigma_{j=1,j\neq i}^{N} r_{ij}^{-1})/N^2$. The results, $R_h$
vs.~$C_s$, for different chain stiffness $k_a$ are presented in
Fig.~\ref{f:RhvCs}.

We can see that the curves show different variations against $C_s$, depending
on the chain stiffness. These variations can be classified into three types.
The first type is attributed to the case of small $k_a$ such as $k_a=0$.  The
chain is said \textit{flexible} and $R_h$ in the semi-log plot looks similar to
a tilted-L curve.  The second type is addressed to intermediate $k_a$, such as
$k_a=8$.  The chain is called \textit{semiflexible}, for which the $R_h$ curve
takes a shape similar to the letter 'U'. The third type is ascribed to large
$k_a$ like $k_a=100$. The chain is very stiff and there is basically no
variation against $C_s$; we are in the case of \textit{rigid} chains.  

Snapshots for these three types of variations are shown by the side of the
figure. For a flexible chain (snapshots (i) to (iii)), the chain size
decreases gradually as $C_s$ increases up to $C_s^*$; the chain undergoes a
continuous coil-to-globule transition. The chain size then turns to increase
slowly when $C_s>C_s^*$, indicating a swelling transition.  $C_s^*$ is found
equal to $8\times 10^{-5}$, the equivalence point. At this particular salt
concentration, the added tetravalent counterions are in charge equivalence with
the bare charge of the PE chain. We observed that the chain displays a
disordered, globule structure at $C_s^*$. For a semiflexible chain (snapshots
(iv) to (vi)), two sharp transitions take place at low and at high salt
concentrations, respectively.  They are, in turn, the coil-to-globule and the
globule-to-coil transitions. It has been shown that these two transitions occur
in a discrete manner~\cite{wei07}, in contrast to the continuous transitions
happened for a flexible chain. In the region between the two salt
concentrations, the chain is in a collapsed state and exhibits a compact
ordered structure, such as a toroidal structure or a folded-chain structure.
Outside the salt region, the chain structure becomes extended. This kind of
phenomena is called ``reentrant condensation'' and has attracted much attention
in the study of DNA condensation~\cite{bloomfield96,bloomfield97}. For the case
of a rigid chain (snapshot (vii)), the chain is so stiff that it cannot be
folded by the condensed multivalent counterions. The chain stays elongated, no
matter how many salts are added in the solution.

\begin{figure} \centering
\includegraphics[height=0.7\textwidth,angle=270]{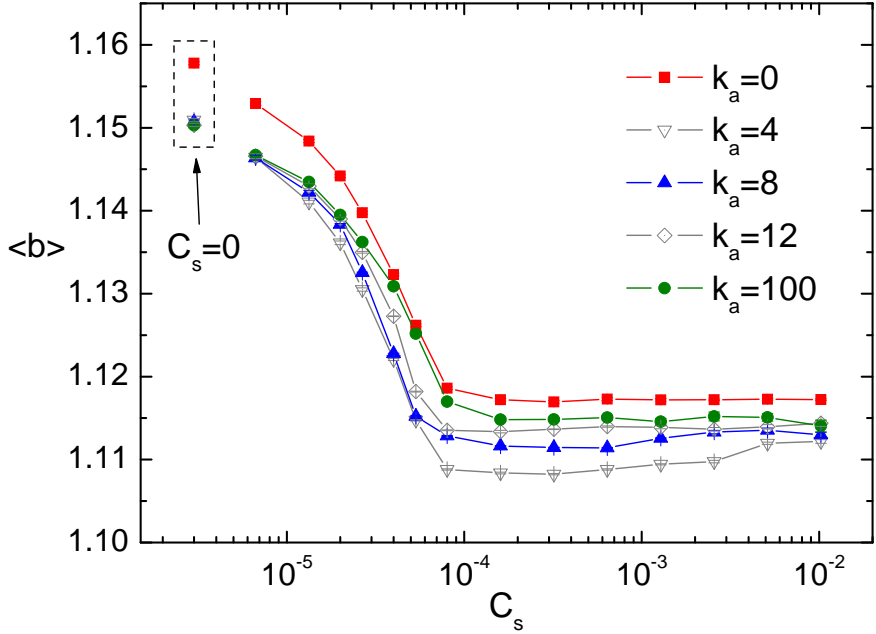} \caption{ $\langle
b \rangle$ as a function of $C_s$ for different chain stiffness $k_a$.  The
values of $k_a$ are indicated in the figure. The data for $C_s=0$ are also
plotted in the dashed box on the left-top side of the figure.} \label{f:blvCs}
\end{figure}

In addition to the chain size and morphology, which are global properties of
chain in solutions, we studied also local properties such as the average bond
length. This information tells us how the bond length changes in different
conditions, which is directly related to the line charge density of a PE chain,
and is important, later on, in the comparison with the Manning condensation
theory. The average bond length $\left<b\right>$ as a function of $C_s$ is
shown in Fig.~\ref{f:blvCs}. 
The chains of different stiffness show similar variational behavior.
$\left<b\right>$ decreases with increasing $C_s$ when $C_s \le C_s^*$. It then
stays basically at a constant for $C_s>C_s^*$. The maximum reduction of
$\left<b\right>$ is less than $4\%$. We remark that this decreasing in
$\left<b\right>$ is a result of the condensation of multivalent counterions on
a PE chain, which form triplets with monomers (as we will see in Section
\ref{subsec_rdf}) and hence pull closer the distance between two neighboring
monomers via the Coulomb interaction. In the region $C_s>C_s^*$, although the
chains are charge overcompensated (cf. Fig.~\ref{f:tube_rc3}), the extra number
of the condensed ions cannot pull closer the monomer-monomer distance anymore
due to the excluded volume effect.  $\left<b\right>$ is thus saturated at a
constant value. Combining with the information that $R_h$ increases when
$C_s>C_s^*$, we know that the bond length is not responsible for the global
chain reexpansion in this region.

\subsection{Ion distribution in chain radial direction}
\label{subsec_ion_dist_in_chain_radial_direction}
\begin{figure} \centering
(a)\includegraphics[height=0.5\textwidth,angle=270]{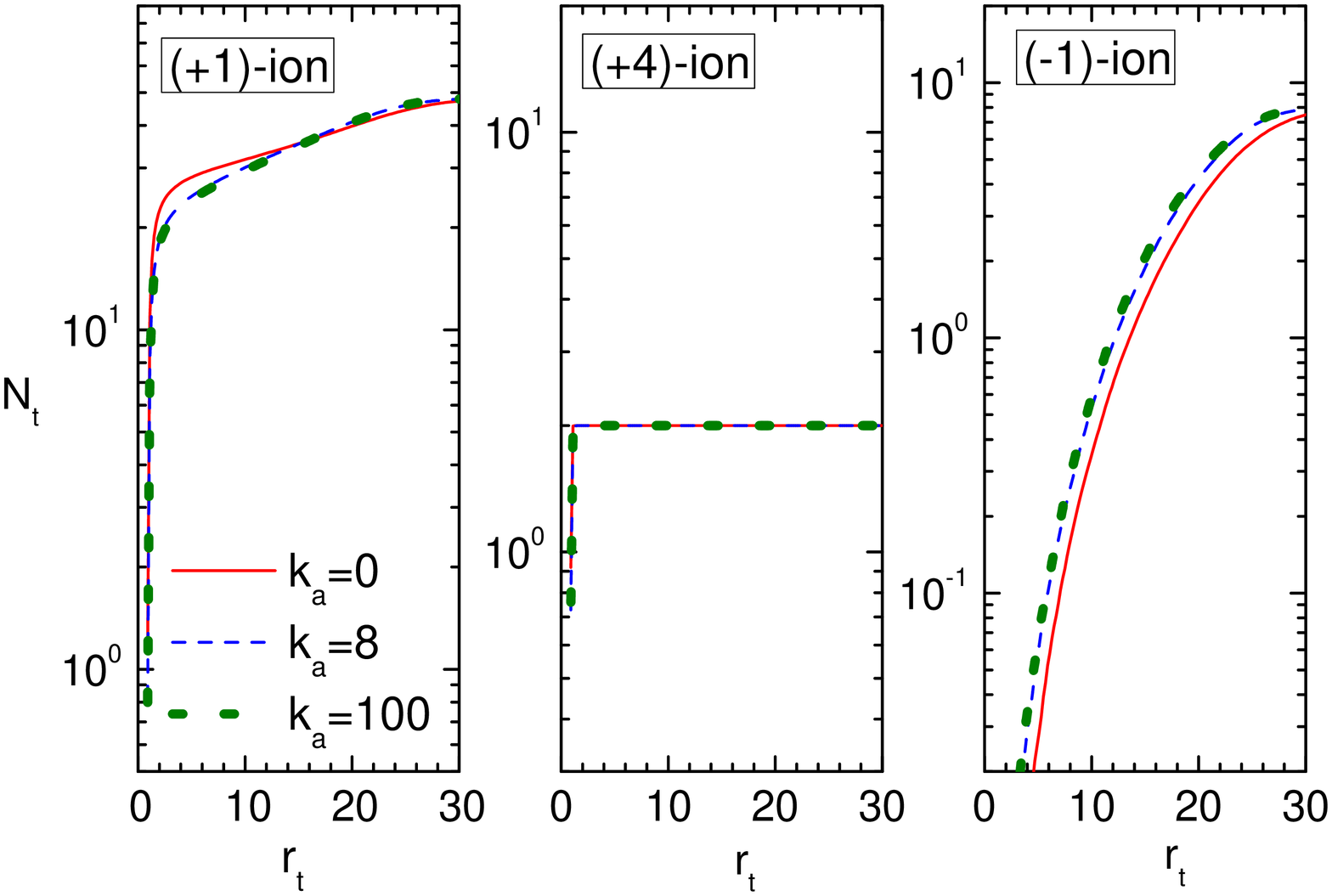}
(b)\includegraphics[height=0.5\textwidth,angle=270]{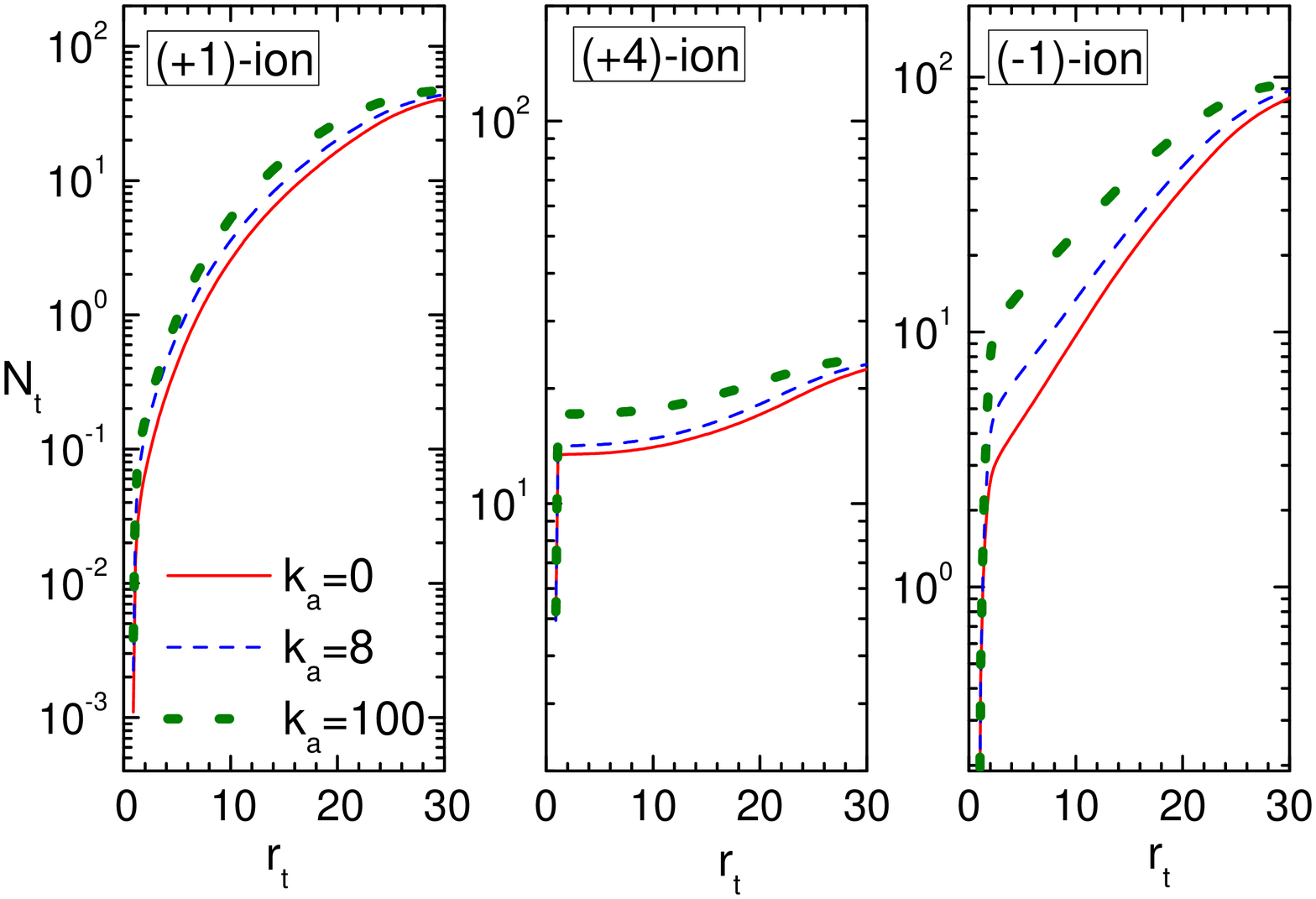}
(c)\includegraphics[height=0.5\textwidth,angle=270]{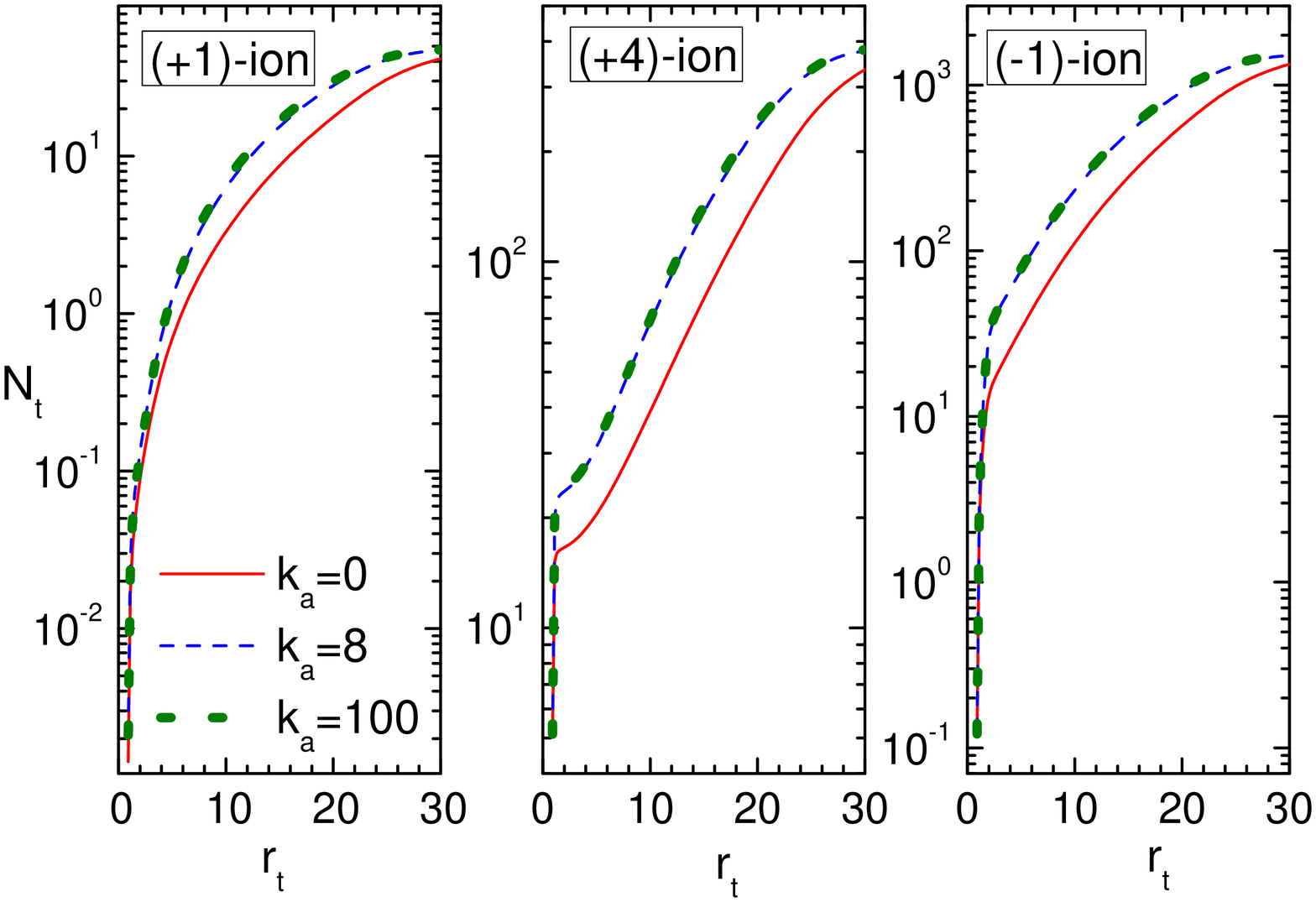}
\caption{Integrated number of ions $N_t(r_t)$ for monovalent counterions
((+1)-ion), tetravalent counterions ((+4)-ion), and coions ((-1)-ion), around a chain 
at $C_s=1.33 \times 10^{-5}$ (a),  $1.6 \times 10^{-4}$ (b), and $ 2.56
\times 10^{-3}$ (c).  The values of chain stiffness $k_a$ are indicated in the
figure. } \label{f:iondist3} \end{figure}

After knowing the morphology of PEs in different salt and chain stiffness
conditions, we study now different kinds of ion distributions and their
relation with the chain morphology. We focus firstly on the ion distribution
in chain radial direction.  A tube region around a chain is defined as the
union of spheres of radius $r_t$, centered at each monomer center. The average
number of ions inside this tube region is calculated for each ion species. By
varying $r_t$, we are able to investigate the integrated ion distribution
$N_t(r_t)$  around the chain radially. The results for monovalent counterions,
tetravalent counterions, and coions  are shown in Fig.~\ref{f:iondist3}.

Fig.~\ref{f:iondist3}(a) shows typical ion distributions in the low salt
region, $C_s<C_s^*$. Manning condensation theory~\cite{manning69,oosawa71}
states that in salt-free solutions, if the value of $\xi$, defined as the ratio
of the Bjerrum length $\lambda_B$ to the charge distance on the chain backbone
$\left< b\right>$, is greater than unity, the condensed counterions will
neutralize the fraction $1-\xi^{-1}$ of charge on the rigid PE chain.
According to this theory, there should be about $61.7\%$ of the monovalent
counterions, or equivalently 30 monovalent counterions, condensing on the PE
chain at $C_s=0.0$ in our simulation condition ($\lambda_B=3$ and $\left<
b\right> \simeq 1.15$; thus, $\xi=2.61$). In this figure, we see only
about 20 monovalent counterions condensing on the chain (inside the tube region
with $r_t=\lambda_B=3$). This is because the tetravalent counterions added in
the solution compete with the monovalent ones and replace part of the condensed
monovalent counterions on the chain.  Therefore, the number of condensed
monovalent counterions is smaller than that in the sat-free solution.
Moreover, we observed that when $r_t>3$, the integrated number $N_t(r_t)$ for
tetravalent counterions is a constant. This result indicates that nearly all
the tetravalent counterions condense on the chain in this low salt region.
Fig.~\ref{f:iondist3}, panels (b) and (c), show typical ion distributions in
the middle ($C_s \sim C_s^*$) and in the high ($C_s>C_s^*$) salt regions.  We
saw that $N_t(r_t)$ for monovalent counterions is almost zero inside the tube
$r_t=3$.  Thus, the condensed ions are now tetravalent in majority. For
tetravalent counterions, $N_t(r_t)$ is no more a constant when $r_t>3$ . There
are hence non-condensed tetravalent counterions presented in the bulk solution.

Form the above three figures, we can see that the ion distributions depend
strongly on the chain morphology.  For example, at low $C_s$
(Fig.~\ref{f:iondist3}(a)), the morphology of the flexible chain ($k_a=0$) is
an extended coil, whereas the semiflexible chain ($k_a=8$) is elongated and
rodlike, resembling the rigid chain ($k_a=100$).  Therefore, the ion
distribution curves are nearly identical for the two cases $k_a=8$ and
$k_a=100$.  Similar results are found at high $C_s$ (Fig.~\ref{f:iondist3}(c))
where the chains reenter into the solution.  In that case, the reentered
semiflexible chains is as extended as a rod, different from the coil structure
of a flexible chain. As a consequence, the $N_t(r_t)$ curve is distinguishable
to that for the flexible chain, but similar to the rigid chain.  In the
mid-salt region (Fig.~\ref{f:iondist3}(b)), the flexible and the semiflexible
chains are collapsed, due to the condensation of tetravalent counterions, but
the collapsed chain structures are different: a disordered globule structure
for the former chains and an ordered structure for the latter.  $N_t(r_t)$
curves are therefore non-identical for the three chain stiffness.  

\begin{figure} \centering
(a)\includegraphics[height=0.5\textwidth,angle=270]{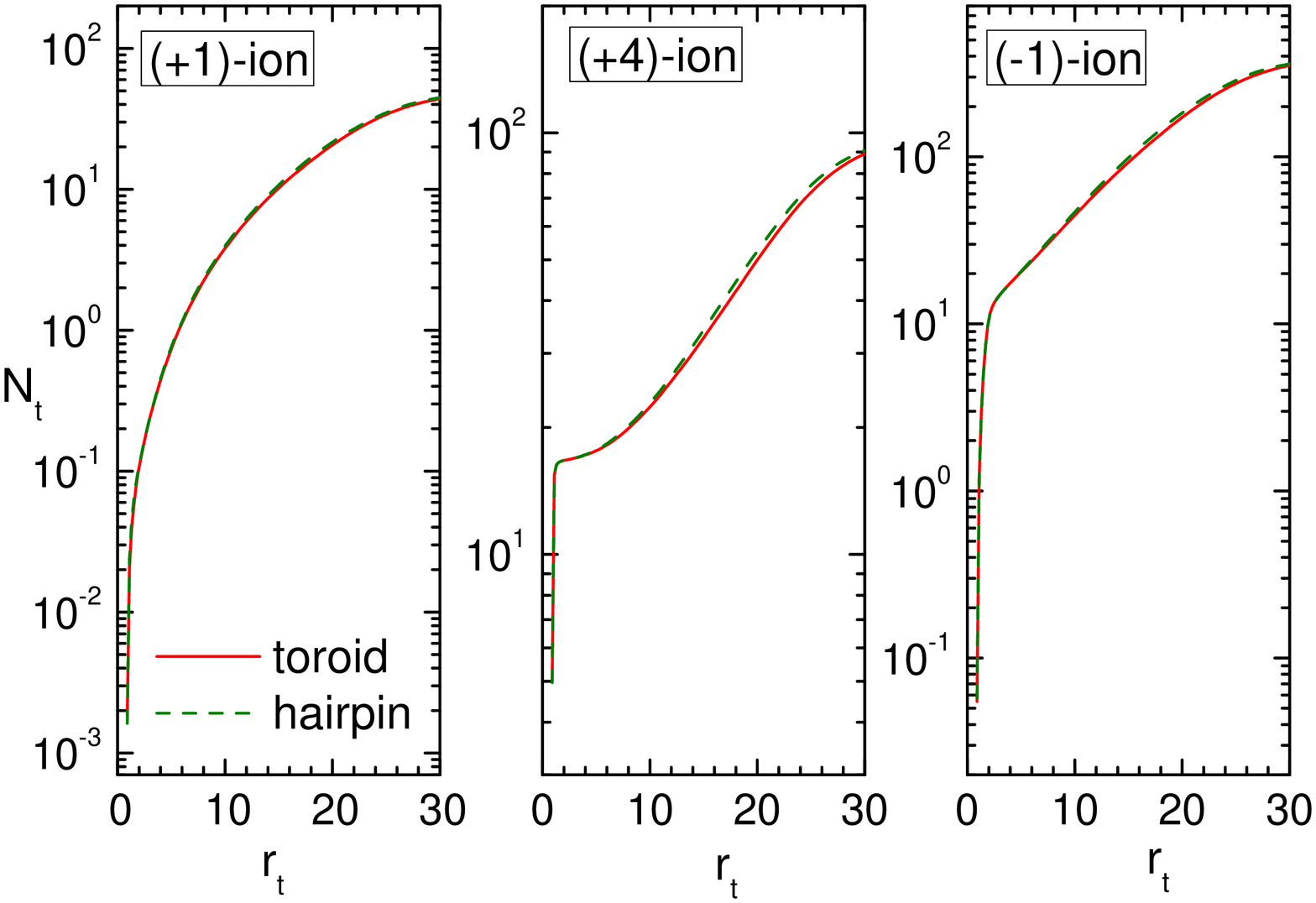}
(b)\includegraphics[height=0.5\textwidth,angle=270]{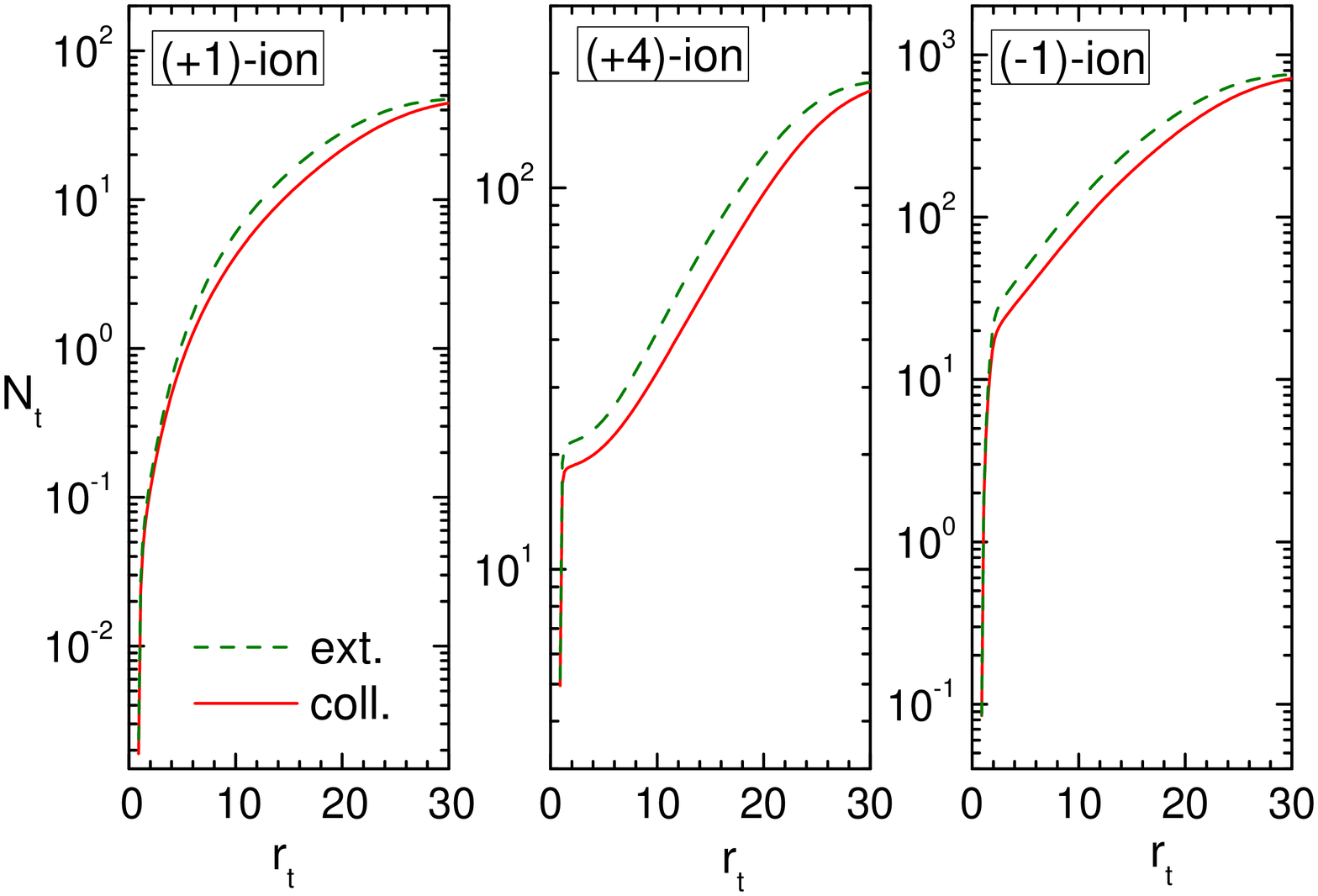}
\caption{$N_t(r_t)$ for the semiflexible chains ($k_a=8$)
(a) at $C_s=6.4\times 10^{-4}$ where the chains are of 
toroid and of hairpin structures, (b) at $C_s= 1.28\times 10^{-3}$ 
where the chains are in an extended (ext.) state and in a collapsed (coll.) state.} 
\label{f:iondist-semi} \end{figure}

The ion distributions for the semiflexible chain ($k_a=8$) were further
investigated in detail in the middle salt region because in this region the
chains can collapse into one of the two typical structures: toroid and hairpin
structures.  We observed that (in Fig.~\ref{f:iondist-semi}(a)) toroidal chains
and hairpin chains show similar ion distributions.
The reason is that the condensed tetravalent counterions bridge between chain
segments in these two cases to force the chains to form toroid or hairpin.
Consequently, the ion distributions look similar along the viewpoint of the chain
radial direction. Moreover, when the salt concentration is close to the
globule-to-coil transition, our previous study showed that the semiflexible
chains can alter between three structures: the toroid, the hairpin and the
extended-rod structures~\cite{wei07}. Since the toroid and the hairpin
structures give a similar distribution of ions around a chain, it is relevant
to verify whether or not the $N_t(r_t)$ curve for the extended-rod chain
structure is also distinguishable to the collapsed chain structures (toroid and
hairpin).  The results are presented in Fig.~\ref{f:iondist-semi}(b) and
significant differences are observed. We found that the more the chain morphology
is opened or extended, the larger the $N_t(r_t)$ will be. This is because more
space is available for ions to condense or to approach to a chain when the
chain has an opened structure. For example, in Fig.~\ref{f:iondist3}(b), the
elongated chains ($k_a=100$) have a larger value of $N_t(r_t)$ than the toroid
or hairpin chains ($k_a=8$), and than the chains collapsed into disordered
globule structure ($k_a=0$).  Similar results have been also observed in
Figs.~\ref{f:iondist3}(a), \ref{f:iondist3}(c) and \ref{f:iondist-semi}(b).
The above evidences show that ion distributions around a chain are determined
by chain morphology, while chain morphology depends on chain stiffness and
salt concentration.      

\subsection{Charge distribution in chain radial direction}
\label{subsec_charge_dist_in_chain_radial_direction}
\begin{figure} \centering
\includegraphics[height=0.7\textwidth,angle=270]{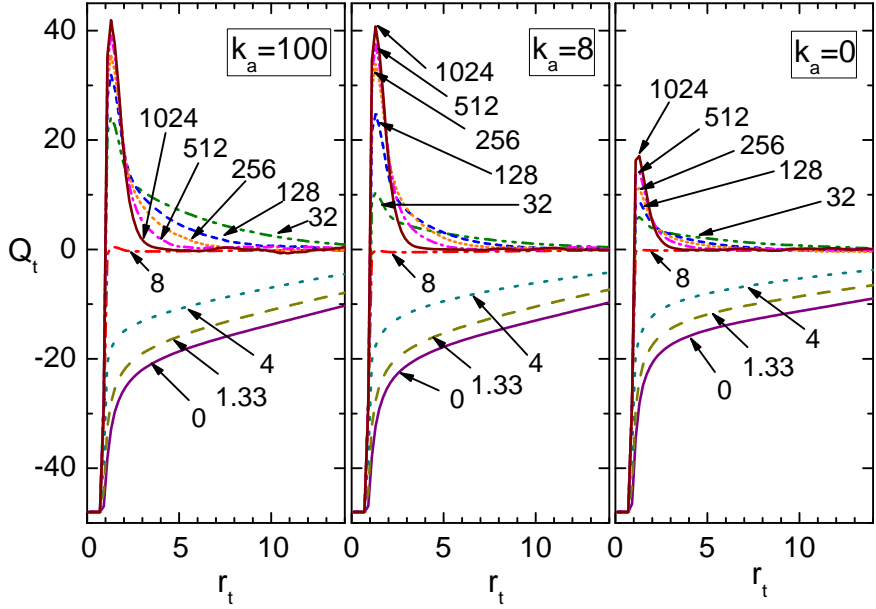} \caption{$Q_t(r_t)$
for the rigid ($k_a=100$), the semiflexible ($k_a=8$), and the flexible
($k_a=0$) chains. The number ($\times 10^{-5}$) near each curve is the salt
concentration $C_s$.} \label{f:Q_m4+} \end{figure}

Using the information obtained in the previous section, by summing up all of
the charges inside a tube region around a chain, we studied the integrated
charge distribution $Q_t(r_t)$.  The results are shown in Fig.~\ref{f:Q_m4+}
for the flexible, the semiflexible, and the rigid chains, over a broad range of
salt concentration.
We observed that when $C_s <C_s^*=8 \times 10^{-5}$, $Q_t(r_t)$ is a monotonic
increasing function of $r_t$ with the value increasing from $-48$, the bare
charge of chain, to $0$, due to the electro-neutrality of the system. In this
salt region, the value of $Q_t(r_t)$ increases with $C_s$ at a given $r_t$.
When the system is at the equivalence point $C_s^*$, the curve $Q_t(r_t)$ turns
out to be zero outside the region of chain excluded volume ($r_t>1$).  This
result tells us an important information that the condensed ions completely
neutralize the chains such that the total charge inside the tube region is
zero.  When $C_s > C_s^*$, $Q_t(r_t)$ starts to show a positive peak near the
chain surface at $r_t \simeq 1.2$. The peak decays rapidly to zero.  The result
indicates the occurrence of charge overcompensation; that is, there is an
excessive number of tetravalent counterions condensing on the chain surface
than needed to neutralize it. A charge overshooting thus takes place. The
higher the salt concentration, the more pronounced the peak will be. Moreover,
we found that the rigid and the semiflexible chains are more overcharged than
the flexible chain.  This is because the opened chain structure provides larger
space to condense multivalent counterions, leading to a more pronounced
$Q_t(r_t)$ peak.

\subsection{Ion condensation on a chain and the effective chain charge}
\label{subsec_ion_condensation_&_effective_charge}
\begin{figure} \centering
(a)\includegraphics[height=0.45\textwidth,angle=270]{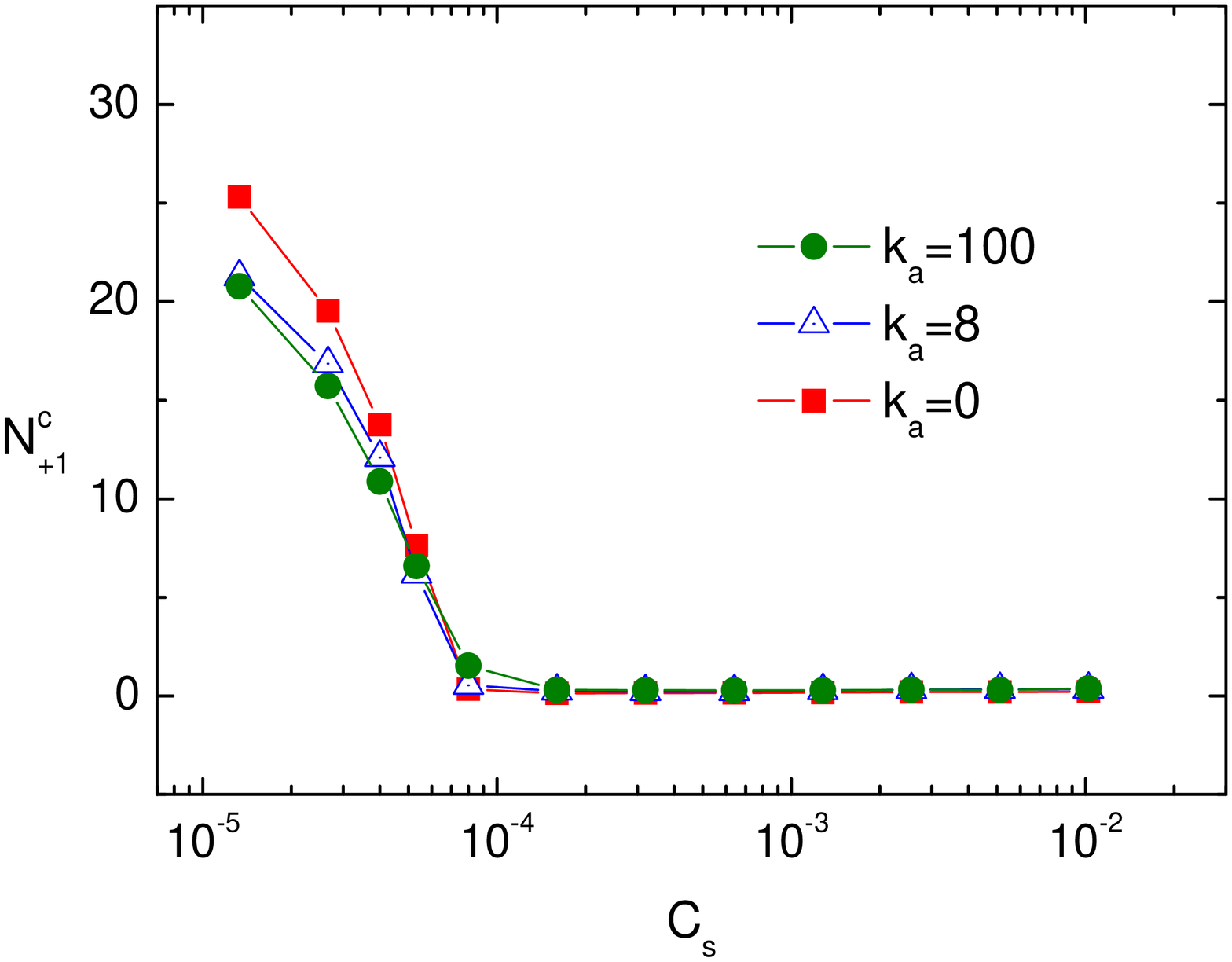}
(b)\includegraphics[height=0.45\textwidth,angle=270]{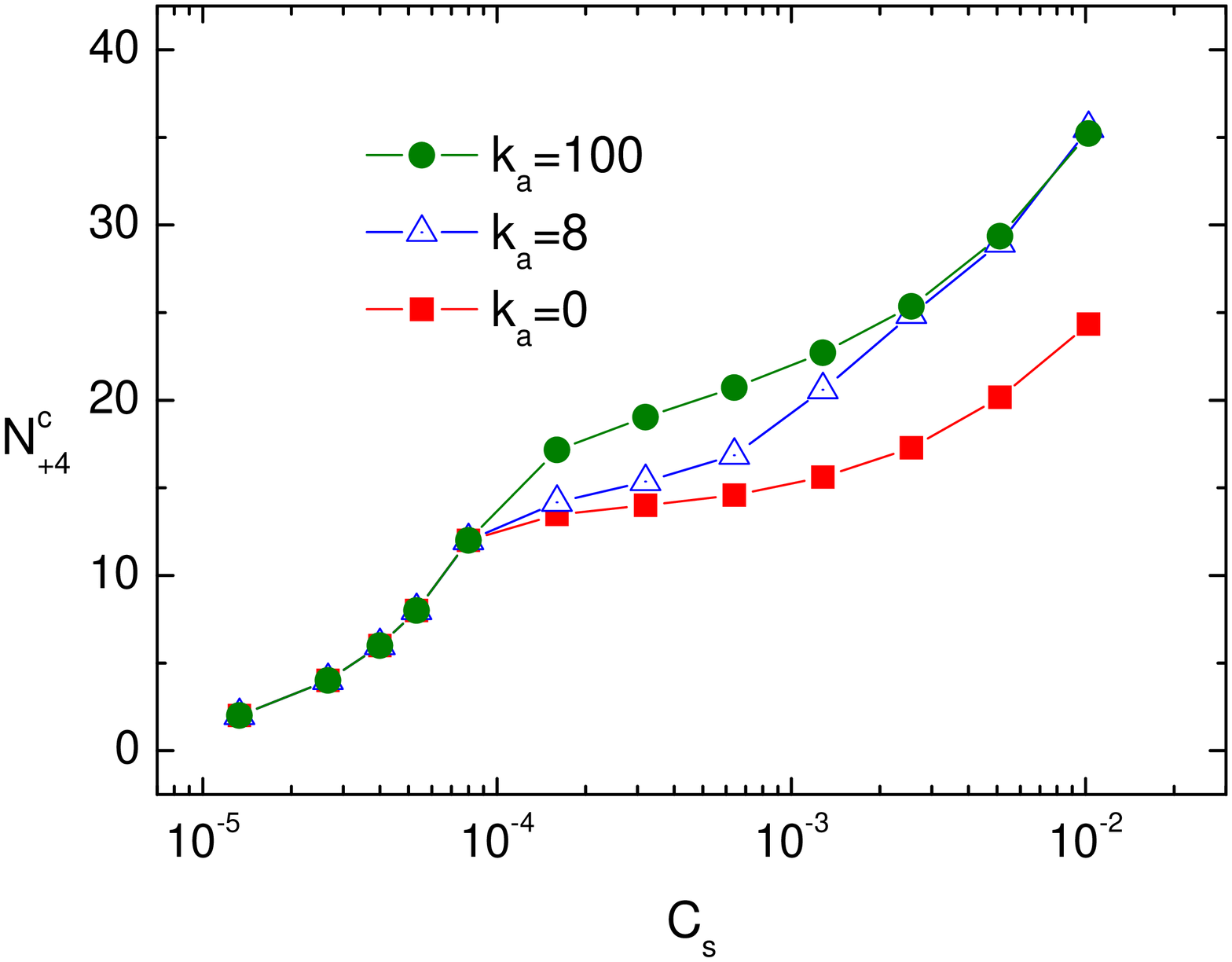}
(c)\includegraphics[height=0.45\textwidth,angle=270]{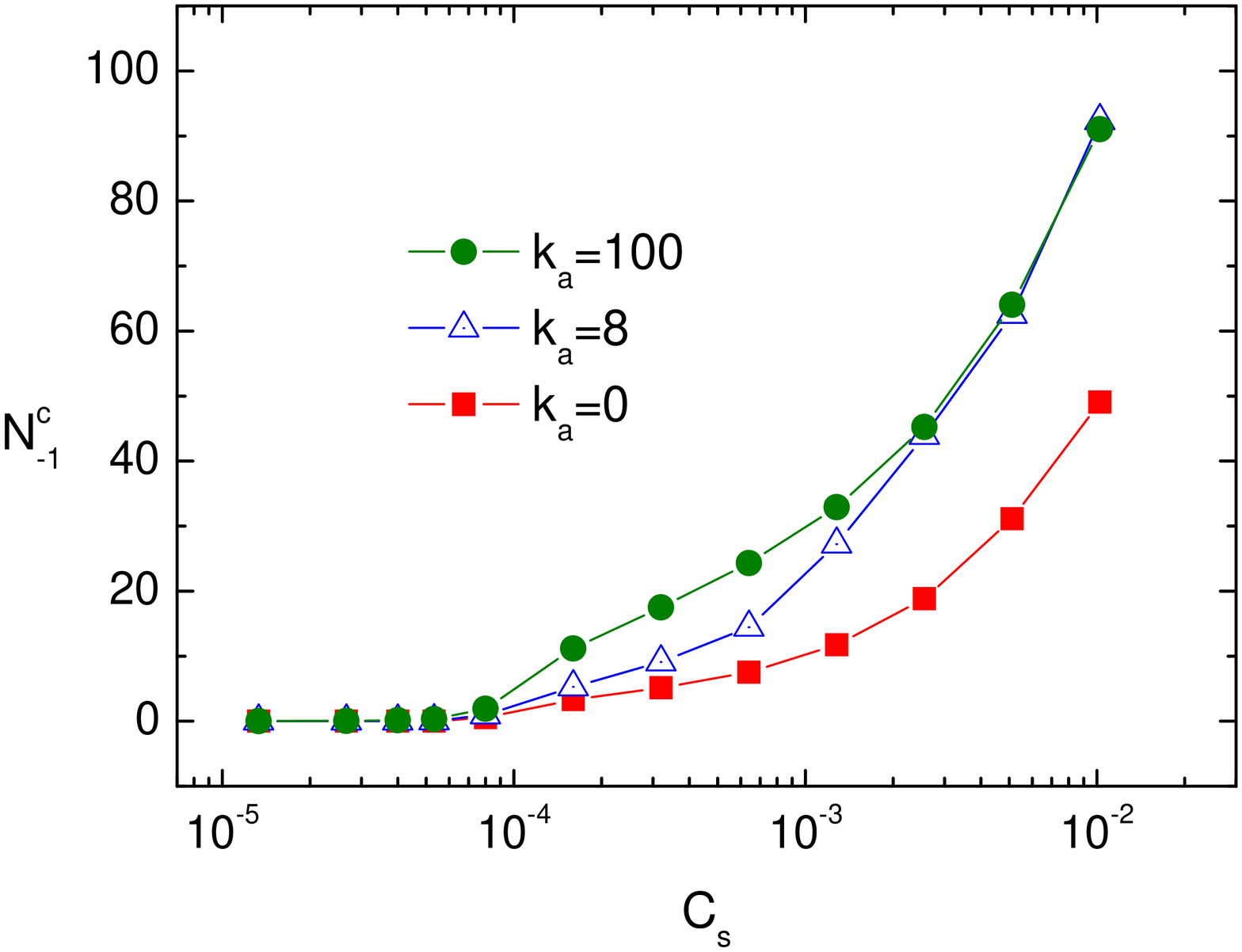}
(d)\includegraphics[height=0.45\textwidth,angle=270]{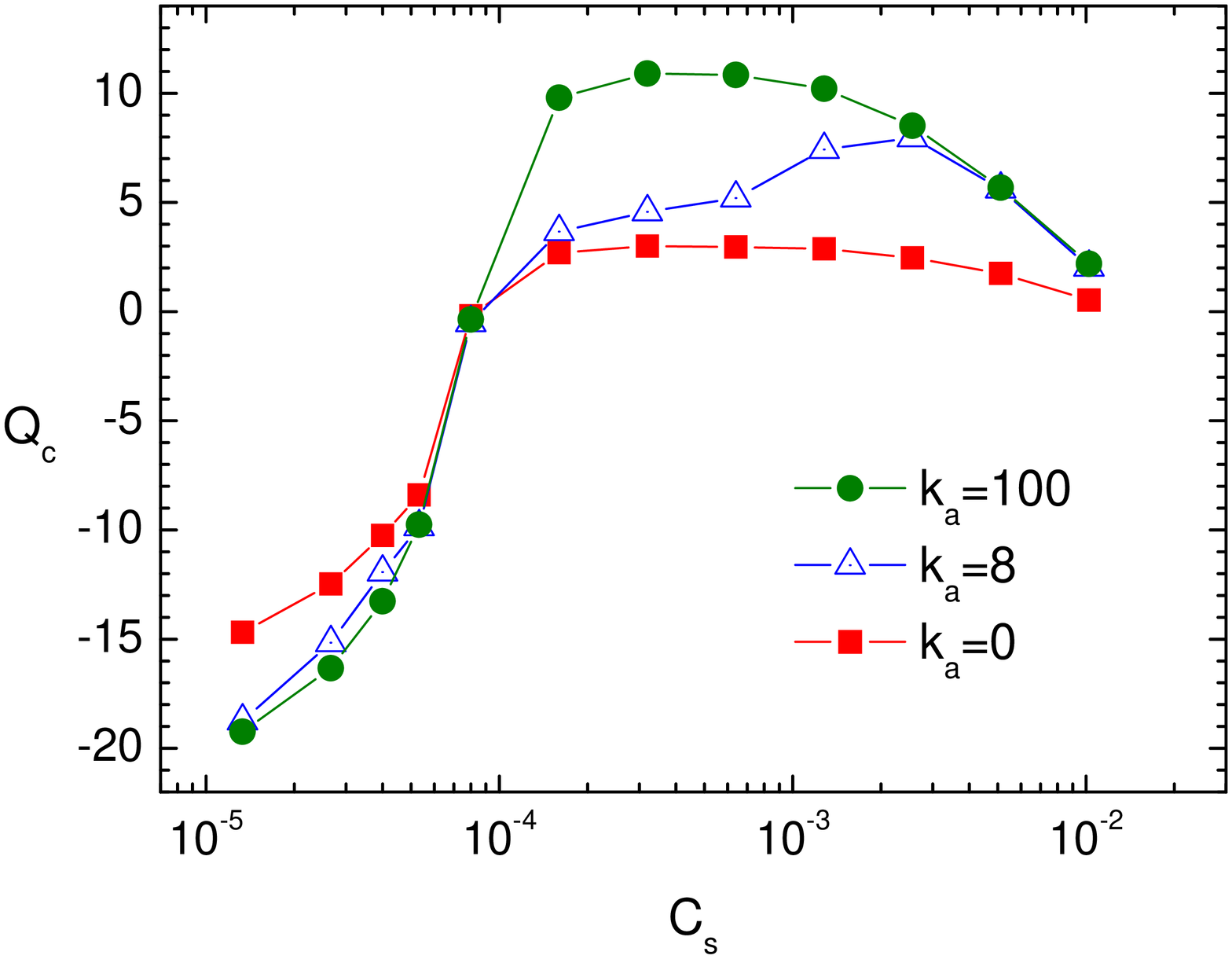}  \caption{(a)
$N_{+1}^c$, (b) $N_{+4}^c$, (c) $N_{-1}^c$, and (d) $Q_c$ as a function of
$C_s$. The values of chain stiffness $k_a$ are indicated in the figures.}
\label{f:tube_rc3} \end{figure}

In the last two sections, the integrated ion and charge distributions were
studied as a function of $r_t$. These figures contain detailed information at
any given tube radius. Nonetheless, in a general discussion, people usually
concern about the number of the condensed ions on a chain and pay more
attention on the effective chain charge resulting from the ion condensation. To
study these topics, one needs to know firstly the region of ion condensation.
However, there exists no clear boundary to separate the condensation region
from the non-condensation one.  We remark that the condensation region is
simply a concept, which can be only qualitatively described.  In this study, we
define the condensation region to be the tube region around a chain with $r_t$
equal to the  Bjerrum length $\lambda_B$ ($=3$).  This definition is based upon
the Manning condensation theory~\cite{manning69,oosawa71}. The results for the
three different chain stiffness are shown in Fig.~\ref{f:tube_rc3}, panels
(a)--(d), where the numbers of the condensed monovalent counterions
($N_{+1}^c$), the condensed tetravalent counterions ($N_{+4}^c$), and the
condensed coions ($N_{-1}^c$), together with the effective charge of chain
inside the condensation region ($Q_c$), are presented as a function of salt
concentration.

Focus firstly on the lowest salt concentration $C_s=1.33 \times 10^{-5}$ in the
figures.  At this $C_s$, 2 tetravalent counterions and 8 monovalent coions are
added into the simulation box (refer to Table~\ref{tab:ion_no}). The results,
$N_{+4}^c= 2$ and $N_{-1}^c=0$, show that the tetravalent counterions entirely
condense on the chain whereas the coions stay all in the bulk solution. Manning
condensation theory predicts that the condensation of counterions on a rigid
chain in a salt-free solution should result in an effective chain charge
density equal to $1/\lambda_B$ (in charge unit)~\cite{manning69,oosawa71}. In
our simulations, $N=48$ monomers constitute a chain and the charge distance on
a chain is $\left< b \right>\simeq 1.15$ (see Fig.~\ref{f:blvCs}). There are
hence $(1-\left< b \right>/\lambda_B)N = (1-\xi^{-1})N$, about 30 monovalent
counterions neutralizing the chain at $C_s=0.0$, according to the theory.
Since 2 condensed tetravalent counterions could repel 8 monovalent counterions
originally condensed on the chain, back into the bulk solution, we expect a
reduction of the number of the condensed monovalent counterion, $N_{+1}^c$, to
a number of about 22.  This is exactly what we observed in
Fig.~\ref{f:tube_rc3}(a) for the rigid ($k_a=100$) and the semiflexible
($k_a=8$) chains. Please notice that $N_{+1}^c$ for the flexible chain
($k_a=0$) is larger than 22. It suggests a stronger ion condensation. 

Focus next on the low salt region $C_s \le C_s^*=8\times 10^{-5}$. We can see
that $N_{+4}^c$ increases with $C_s$. In fact, $N_{+4}^c$ is equal to the
number of tetravalent counterions added into the simulation box. It shows a
full condensation of these ions on the chain. In this salt region, the
$N_{+4}^c$ curves overlap with each other, regardless of the chain stiffness.
For condensed coions, we found $N_{-1}^c=0$. It suggests that the chains repel
the coions. This fact can be examined in Fig.~\ref{f:tube_rc3}(d) where the
effective chain charge $Q_c$ indeed takes a negative value (when $C_s \le
C_s^*$). Therefore, the coion condensation is unfavored because of the Coulomb
repulsion.  Concerning the condensation of monovalent counterions on a chain,
we found that the number decreases with increasing $C_s$ and hits the minimum
value, zero, at $C_s=C_s^*$. At the same moment, $Q_c$ increases and crosses
eventually the zero point.  

Focus thirdly on the high salt region $C_s>C_s^*$. We observed that $N_{+4}^c$
continuously increases with $C_s^*$ but splits into different curves, depending
on the chain stiffness. These splitting curves are upper bounded by the curve
coming from the rigid chain, while lower bounded by the one coming from the
flexible chain. For the semiflexible chain, $N_{+4}^c$ follows at beginning the
lower-bounded curve and then transfers gradually to the upper-bounded curve.
Once reexpanded in the solutions ($C_s>3\times 10^{-3}$), the chain looks
similar to a rigid rod and the curve overlaps with the upper-bounded curve.
Please notice that $N_{+4}^c$ is larger than the number needed to neutralize
the bare chain charge. Charge overcompensation hence takes place and the chain
charge turns to be positive. We note that in this salt region, the tetravalent
counterions are not always condensed on the chain; in other words, fraction of
them stay in the bulk solution. Since the chain charge is effectively positive,
the coions can be attracted toward the chain but not the monovalent
counterions.  For this reason, $N_{-1}^c$ starts to increase with $C_s$ (see
Fig.~\ref{f:tube_rc3}(c)) but $N_{+1}^c$ stays at zero (see
Fig.~\ref{f:tube_rc3}(b)). We observed that $N_{-1}^c$ behaves similarly to
$N_{+4}^c$, which split into curves depending on the chain stiffness. This
behavior is directly associated to that of the $N_{+4}^c$ because coion
condensation necessitates the help of bridging by excessively condensed
tetravalent counterions. For the effective chain charge, we did observe that
$Q_c$ becomes positive when $C_s>C_s^*$.  Nevertheless, the $Q_c$ curve does
not trivially follow the $N_{+4}^c$ curve. It shows convex-down behavior at
some salt concentration. The decrease of $Q_c$ tells us that the condensing
rate for coions exceeds 4 times of that for tetravalent counterions beyond the
salt concentration. Again, $Q_c$ splits into different curves when $C_s>C_s^*$
with the upper bound being the case of the rigid chain and the lower bound
being the case of the flexible chain. $Q_c$ for the semiflexible chain deviates
from the upper bound in the mid-salt regions because drastic change of chain
conformation occurs in this region, very distinguishable to the morphology of
the rigid chain. Finally, a trend was observed when $C_s$ is very high: $Q_c$
tends to zero, no matter of which stiffness the chain is. The charged chains
are effectively neutralized.

The results obtained in this section demonstrate again that it is the chain
morphology deciding the ion condensation and the effective chain charge.
The chain stiffness plays a role, behind curtains, to determine the
chain morphology at a given salt concentration.  

\subsection{Distribution of condensed ions in chain axial direction}
\label{subsec_dist_cond_ion_along_chain}
In the previous three sections, we have investigated the ion distribution and
condensation in the radial direction away from a chain axis. However, the
pertinent information concerning the distribution in the perpendicular
direction, \textit{i.e.}, in the direction of the chain axis, is still missing.
In order to fill up this gap of information, we investigate here how condensed ions
distribute along a chain axis. The distribution is calculated by the following
way.  We firstly index the monomers sequentially, starting by 1, from one end of
the chain to the other. We then assign each condensed ion (inside the condensation
region $r_t=3$) to a monomer, to which from the ion the distance is the
shortest.  The average numbers of the condensed ions assigned to every monomer
are thirdly calculated for each ion species. The results give the ion
distribution along the chain.

\begin{figure} \centering
\includegraphics[height=0.7\textwidth,angle=270]{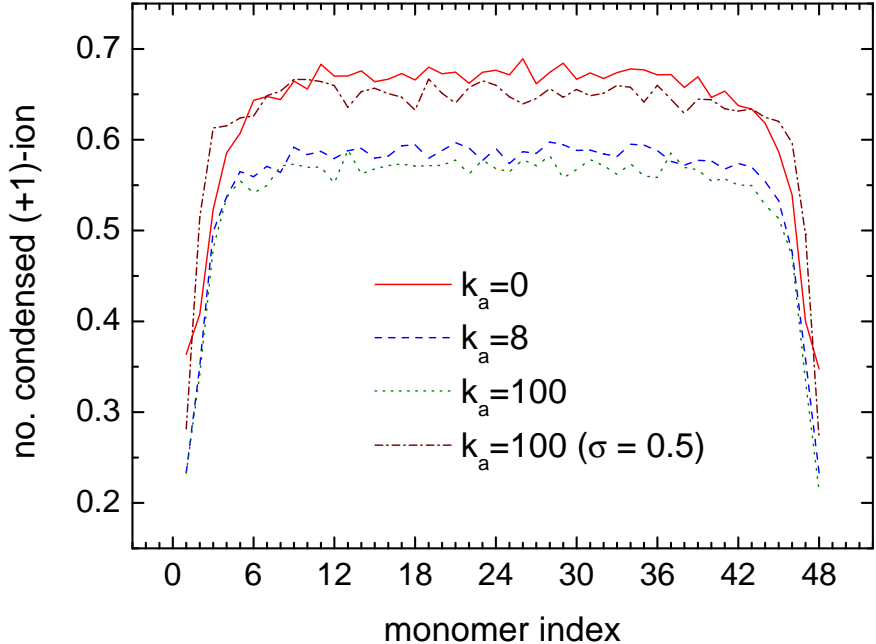} \caption{Mean
number of condensed monovalent counterions vs.~monomer index in a salt-free
solution for three chain stiffness, $k_a=0$, $8$ and $100$. 
In order to show the importance of ion excluded volume, we plotted, in addition, 
the case of $k_a=100$ with the ion size being reduced
to half of the original one, marked by $\sigma=0.5$. }
\label{f:ionpos_S0} \end{figure}

Fig.~\ref{f:ionpos_S0} shows the distribution of the condensed monovalent
counterions in a salt free solution. Each curve denotes one chain stiffness. 
We observed that the distributions show plateau structure in the central part
of the chains, while exhibiting a rapid decrease when the monomers approach to
the two chain ends.  It shows that the counterions are more likely to condense
on the central part than on the two ends.  When condensing onto the central
part, a counterion is attracted by the monomers from both sides of the chain,
while it is attracted only by one side of the chain if the condensation occurs
near the chain ends. The Coulomb attraction is therefore stronger for the
former case, and as a consequence, the condensed counterions are more numerous
in the chain center. The results are in agreement with the works done by other
groups~\cite{limbach01,liu02,sarraguca06}. Moreover, we found that the value at
the plateau is higher for the flexible chain than for the rigid chain, which
shows that monovalent counterions condense more strongly when the chain is
flexible. We know that a flexible chain displays a zig-zagged structure when
counterions condense onto it and becomes less extensive compared to a
semiflexible chain and a rigid chain~\cite{wei07}.  Consequently, the line
charge density on a flexible chain is effectively higher, which can lead to a
stronger ion condensation. To examine it, we calculated the average angle
$\left<\theta\right>$ between two adjacent bonds for the three cases of chain
stiffness, and the results are $\left<\theta\right>=121.8(5)^{\circ}$,
$160.7(2)^{\circ}$, and $174.3(1)^{\circ}$ for the flexible ($k_a=0$), the
semiflexible ($k_a=8$), and the rigid ($k_a=100$) chains, respectively. The
effective charge distance in the main direction of the zig-zagged chain axis
thus can be estimated by $\left<b_{\rm eff}\right>=\left< b\right>
\sin(\left<\theta\right>/2)$.  To adapt it for the chains with flexibility, We
modify the Manning condensation theory by replacing the charge distance on the
chains with the effective charge distance.  The modified theory predicts a
fraction of charge neutralization on the chains equal to $1-\left<b_{\rm
eff}\right>/\lambda_B$, which yields 0.665, 0.622, and 0.617 condensed
monovalent counterions per monomer for the three studied stiffness,
respectively. Our simulations show consistent trend of behavior with the
theory. Nonetheless, the numbers are slightly smaller than the predictions.
This is because the Manning theory was derived under the assumption of
pointlike ions condensing onto an infinite cylindrical chain. Our modeling ions
have excluded volume and the PE chain has finite chain length; both of the
effects can decrease the number of the condensed ions. To verify it, identical
simulations have been performed for the rigid chain ($k_a=100$) except that the
counterions were set to be half of the original ion size. The result is also
plotted in Fig.~\ref{f:ionpos_S0} in dash-dotted curve (denoted by
$\sigma=0.5$) for comparison. We observed that the number of the condensed ions
increases, which shows the importance of ion excluded volume on ion
condensation. Here, the distance of closest approach between the monomer and
the counterion is also a relevant parameter. The smaller the ion size, the
closer the distance, and therefore, the stronger the condensation. 

We next concentrate on the systems with added tetravalent salt. 
Since the tetravalent counterions play a crucial role in determination of  
chain morphology by substituting the condensed monovalent counterions on a chain, 
we study here how the condensed tetravalent counterions distribute 
on a chain. 

\begin{figure} \centering
(a)\includegraphics[width=0.28\textheight,angle=270]{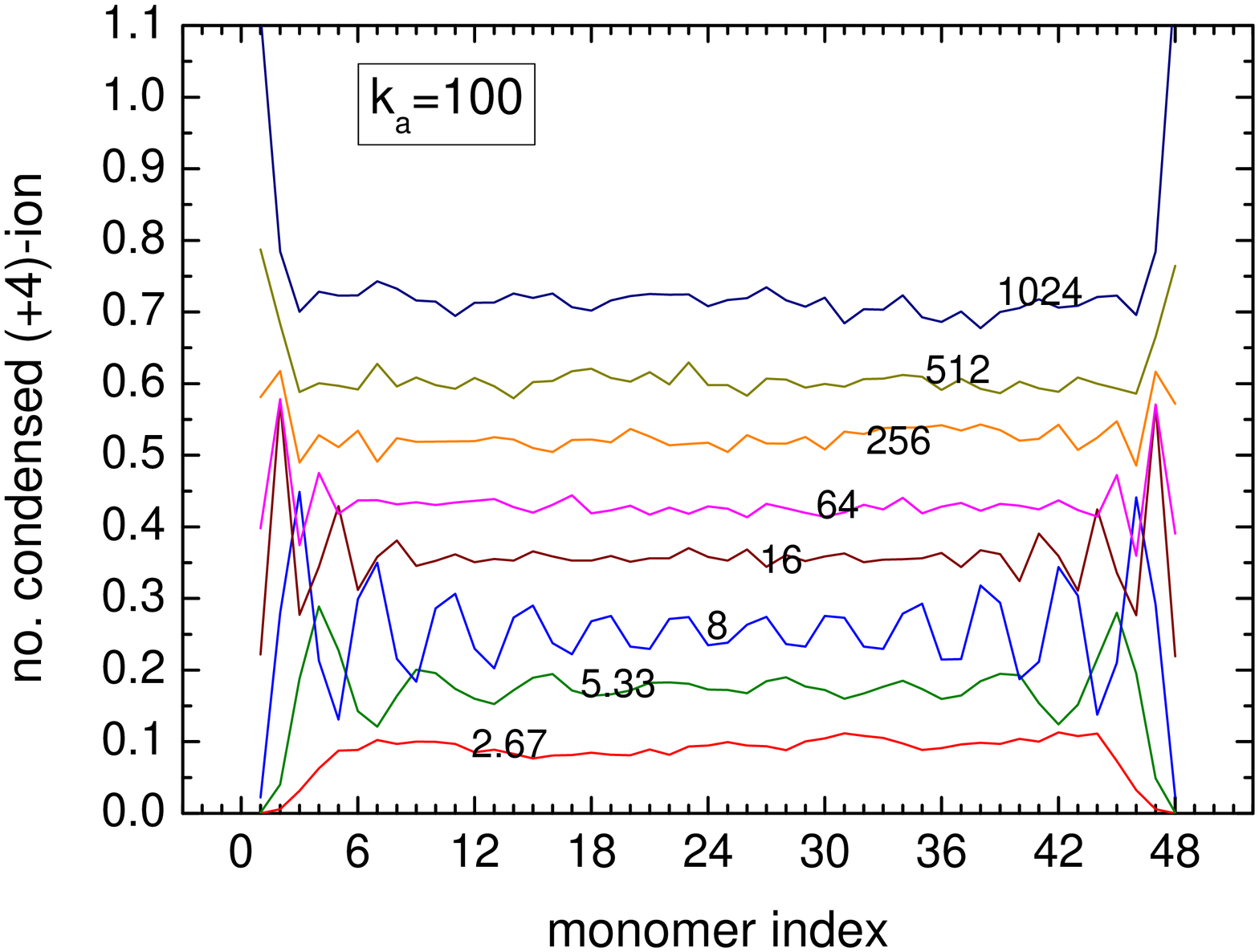}
(b)\includegraphics[width=0.28\textheight,angle=270]{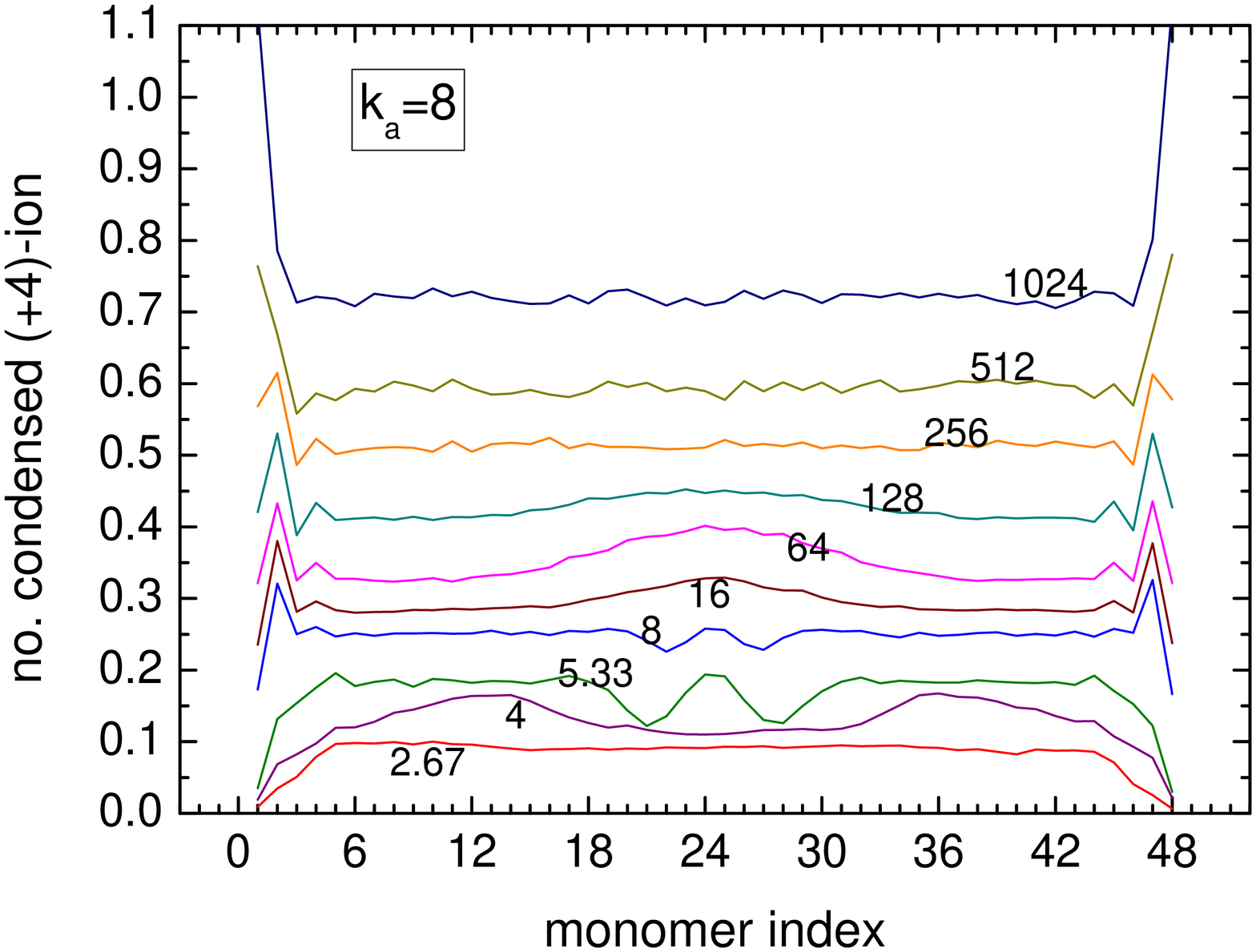}
(c)\includegraphics[width=0.28\textheight,angle=270]{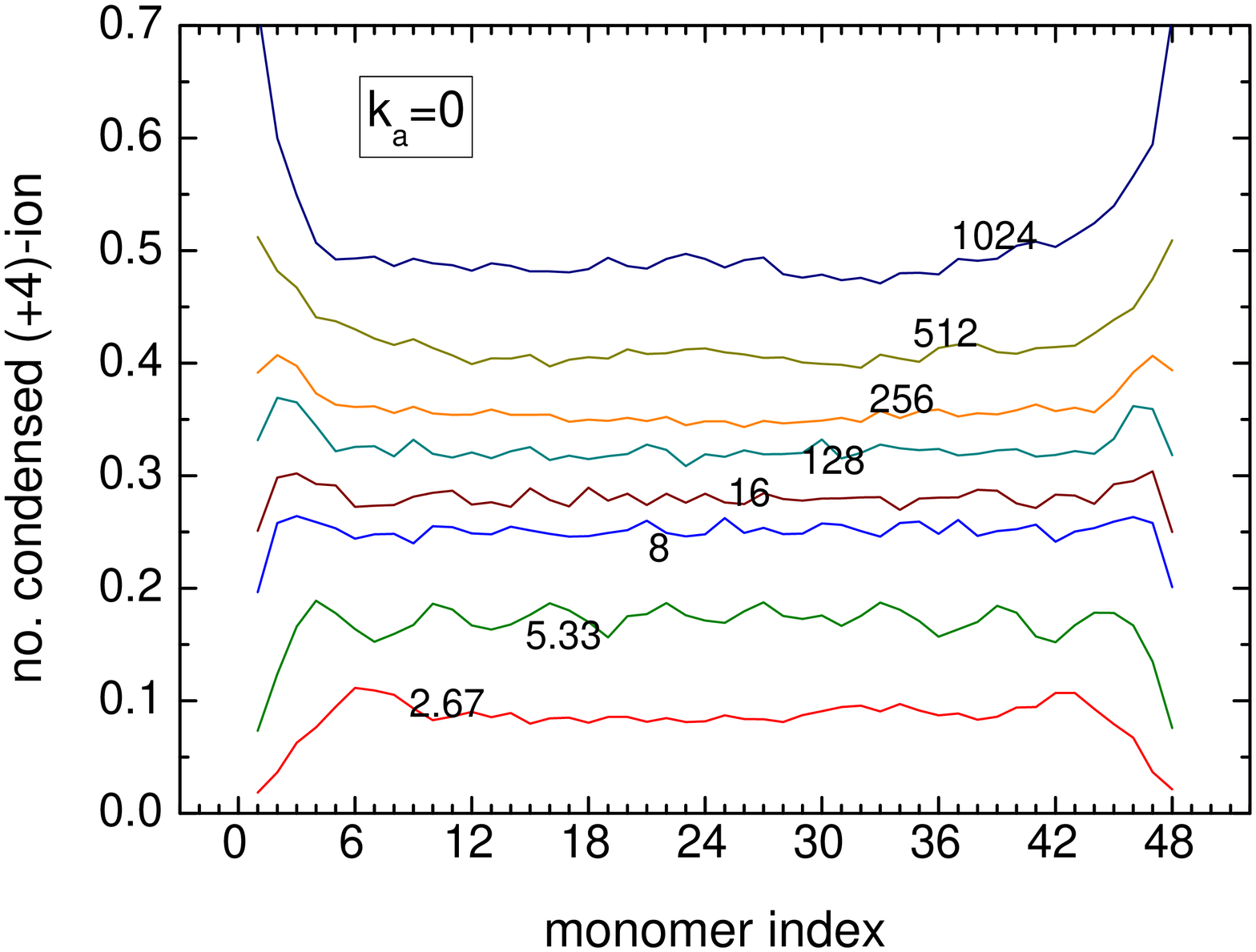}
\caption{Mean number of condensed tetravalent counterions vs.~monomer index for
(a) the rigid chain, (b) the semiflexible chain, and (c) the flexible chain. 
Each curve denotes one distribution at a salt concentration $C_s$.  The value
of $C_s\times 10^{5}$ can be read near the corresponding curve. }
\label{f:ionpos_4+} \end{figure}

Fig.~\ref{f:ionpos_4+}(a) shows such distribution on the rigid chain
($k_a=100$).
At the low salt concentration $C_s=2.67\times 10^{-5}$, the ions distribute
uniformly on the chain backbone except near the chain ends. The distribution
profile looks similarly to that for monovalent counterions in salt-free
solution. This is because many monovalent counterions still condense on the
chain (cf. Fig.~\ref{f:tube_rc3}(a)) and collide with the condensed tetravalent
counterions. The condensed tetravalent ions thus diffuse on the chain,
resulting in a uniform distribution. As $C_s$ increases, the condensation of
tetravalent counterions ensures a rapid decrease of the number of the condensed
monovalent counterions. Eventually, the tetravalent ions become the majority on
the chain and the strong electrostatic repulsion between them induces the
localization of these ions at some specific positions along the one-dimensional
chain. This is why at $C_s=5.33\times 10^{-5}$ and $8.0\times 10^{-5}$, several
peaks appear on the distribution profile. The number of the peaks corresponds
to the number of the tetravalent ions condensed on the chain, which is exactly
the amount of these ions added in the simulation box (cf.
Table~\ref{tab:ion_no}).  When $C_s$ is increased over the equivalence point
$C_s^*(=8.0\times 10^{-5})$, the number of the condensed tetravalent
counterions exceeds the one needed to neutralize the chain backbone charge.
Charge overcompensation takes place. The ion localization is now disturbed by
these extra ions. The disturbing shows its effect starting from the central
part of the chain and propagating to the chain ends with increasing the salt
concentration. Finally, the distribution becomes uniform again. We found that,
different to the cases at low salt concentrations, the tetravalent ions favor
to condense on the chain ends when $C_s$ is high. This is because near the
chain ends, there are more spaces available to accommodate these over-crowded
ions. 

\begin{figure} \begin{center}
\includegraphics[height=0.6\textheight,angle=270]{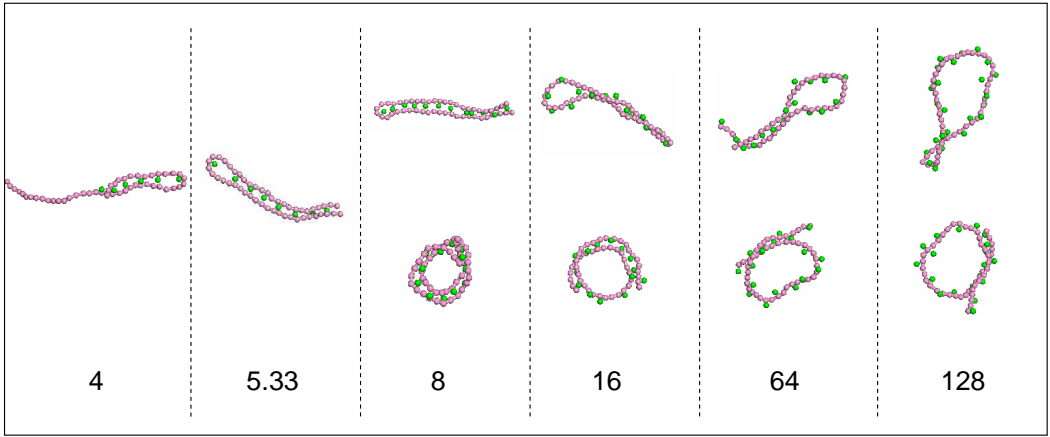}
\caption{Snapshots of typical morphologies of the semiflexible chain ($k_a=8$).
The numbers below the pictures are the salt concentrations ($\times 10^{5}$).
} \label{f:snapshots} \end{center} \end{figure}

Fig.~\ref{f:ionpos_4+}(b) shows the distribution on the semiflexible chain
($k_a=8$). When $C_s<4\times 10^{-5}$ or $C_s>1.28\times 10^{-3}$, the PE chain
displays an extended structure.  Similar to the rigid chain, the condensed
tetravalent ions distribute uniformly on the chain backbone.  In between the
salt concentrations, the chain forms ordered collapsed structure and the
distribution displays some  specific profile. For example, at $C_s=4\times
10^{-5}$, the distribution shows a symmetric profile with two humps. This
profile results from a particular hairpin structure in which the length of the
two branches of the hairpin is not equal (cf.~snapshots in
Fig.~\ref{f:snapshots}). Each hump in the distribution reflects the drastic
change of chain geometry at an eccentric bending point of the hairpin. We note
that the appearance of the two humps do not mean two bendings in one hairpin.
It is a result of ensemble average in which the probability of the bending
point close to one of the two chain ends is equal when the system is in
equilibrium. Therefore, the distribution is a symmetric function with respect
to the middle point.  At $C_s=5.33\times 10^{-5}$ and $8\times 10^{-5}$, the
number of the condensed ions is sufficiently numerous to bind the chain,
against the chain stiffness, to form a hairpin with two equal branches
(cf.~Fig.~\ref{f:snapshots}). A condensed counterion is localized at the
bending point, which results in the peak at the middle point of the chain.  For
$C_s> 8\times 10^{-5}$, the number of the condensed tetravalent ions is excess.
A loop is formed near the hairpin head in order to create more room to
accommodate these extra ions. The chain morphology thus looks similar to a
tennis racquet.  The more the excess number of the ions, the larger the loop
will be formed and the distribution shows a broader and higher hump near the
middle of the chain. We remark that when $8\le C_s\times 10^{5} \le 128$,
toroid is also a favored structure of the chain. The ion distribution on
toroidal structure is similar to the one on a hairpin (or a racquet), which
displays a hump in the middle. To help readers to obtain the whole pictures of
the system in mind, we present in Fig.~\ref{f:snapshots} typical snapshots of
the semiflexible chain together with the condensed tetravalent counterions in
the middle range of salt concentration.

Figure~\ref{f:ionpos_4+}(c) shows the distribution on the flexible PE
($k_a=0$). We have learned that the flexible chain is collapsed into a
disordered structure when the salt concentration is intermediate. The condensed
tetravalent ions cannot be localized at some specific positions of the chain.
The distribution hence shows a uniform distribution. By comparing
Fig.~\ref{f:ionpos_4+}(a), (b) and (c), we noticed that the number of the
condensed tetravalent counterions is less numerous for the flexible chain than
for the semiflexible and the rigid chains when $C_s>C_s^*$. The reason is that
the flexible chain occupies a smaller volume space compared to the other two
chain stiffness (cf.~Fig.~\ref{f:RhvCs}). Consequently, smaller room is
available to accommodate the strongly-repelled tetravalent ions, leading to
fewer condensed tetravalent counterions. It is worth noticing that in the
condensation of the monovalent counterions at $C_s=0$ (cf.
Fig.~\ref{f:ionpos_S0}), the number of condensed monovalent counterions is more
numerous for the flexible chain, which shows an opposite trend to this case.

\subsection{Radial distribution function between monomers and tetravalent
counterions}
\label{subsec_rdf}
\begin{figure} \centering
(a)\includegraphics[width=0.28\textheight,angle=270]{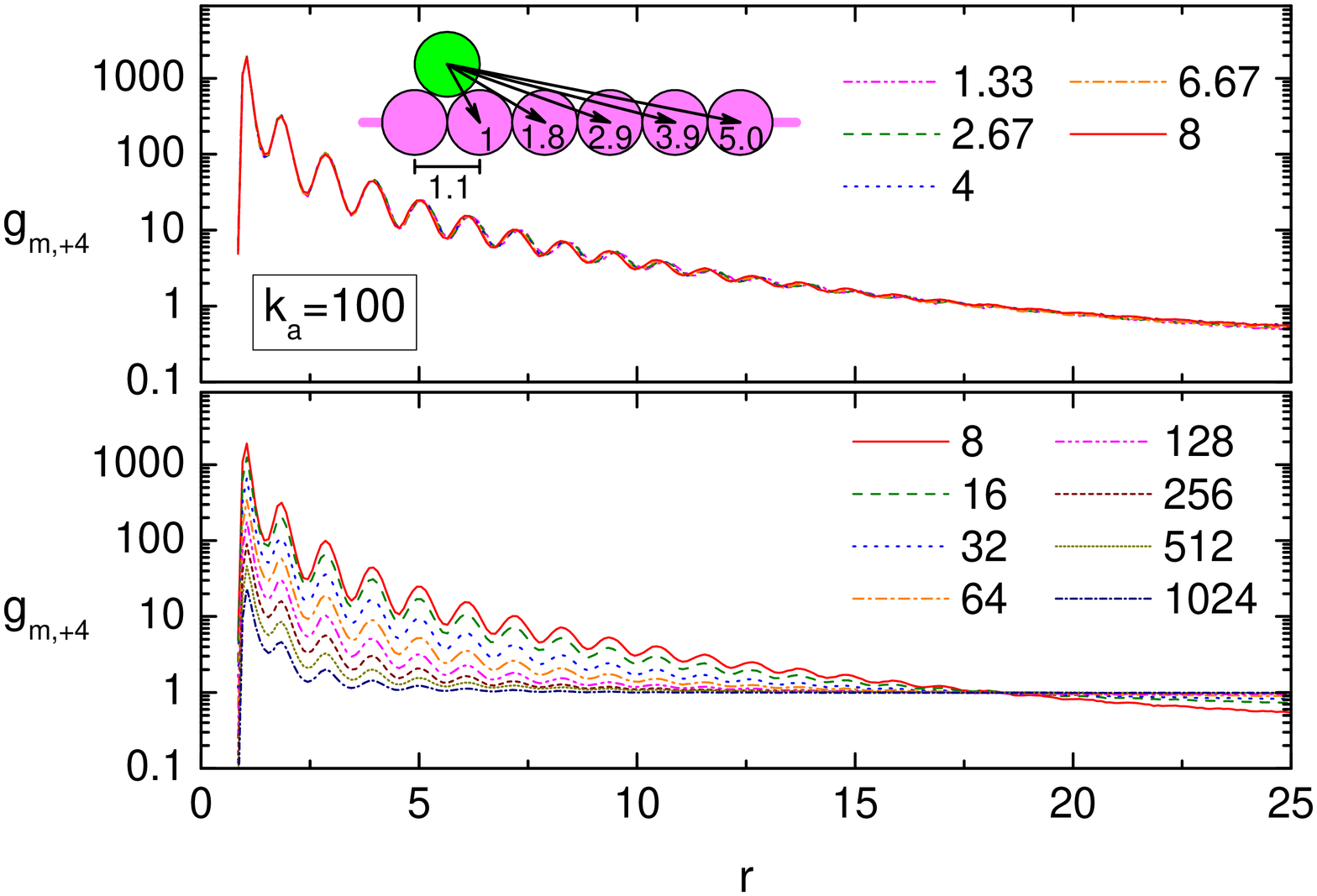}
(b)\includegraphics[width=0.28\textheight,angle=270]{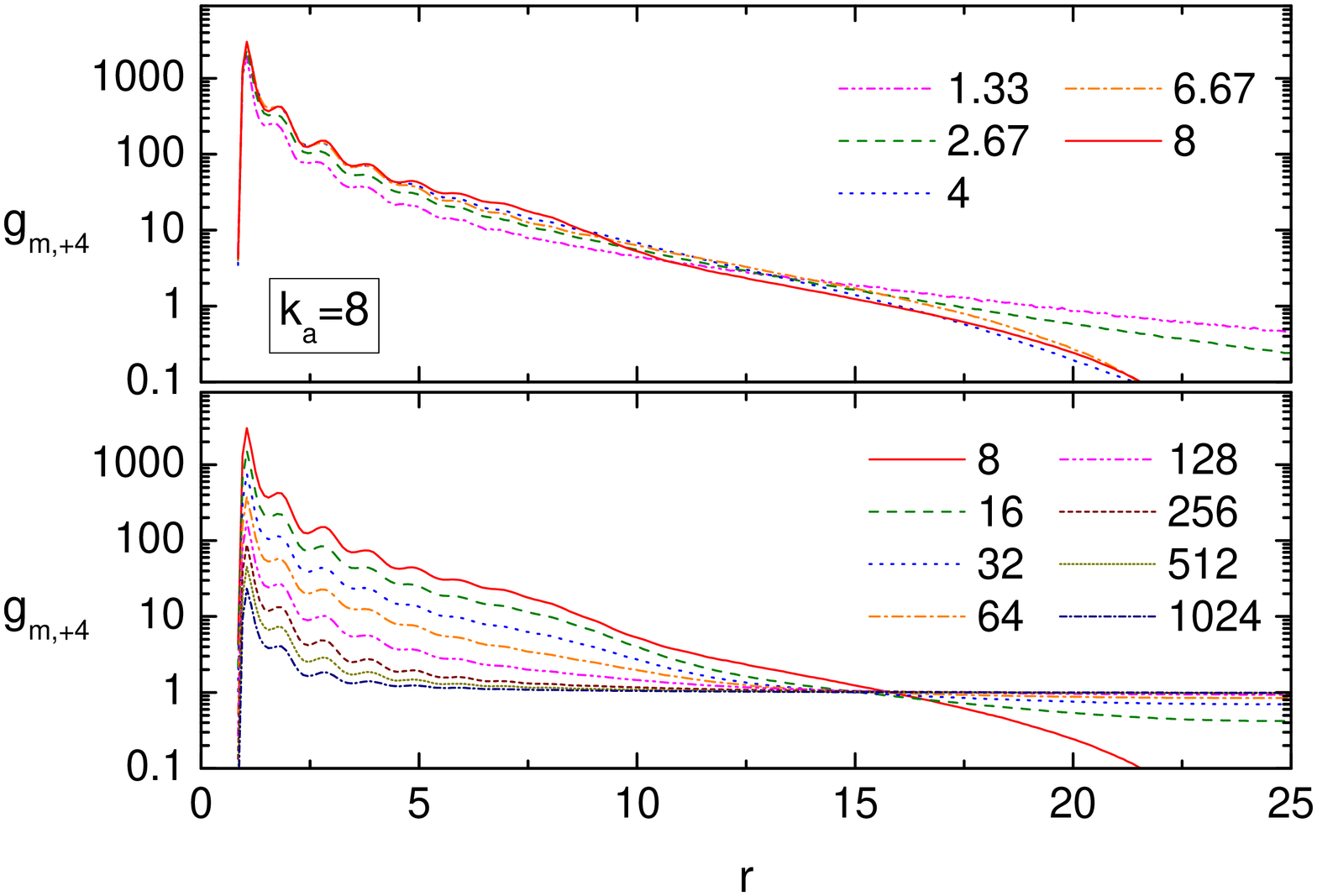}
(c)\includegraphics[width=0.28\textheight,angle=270]{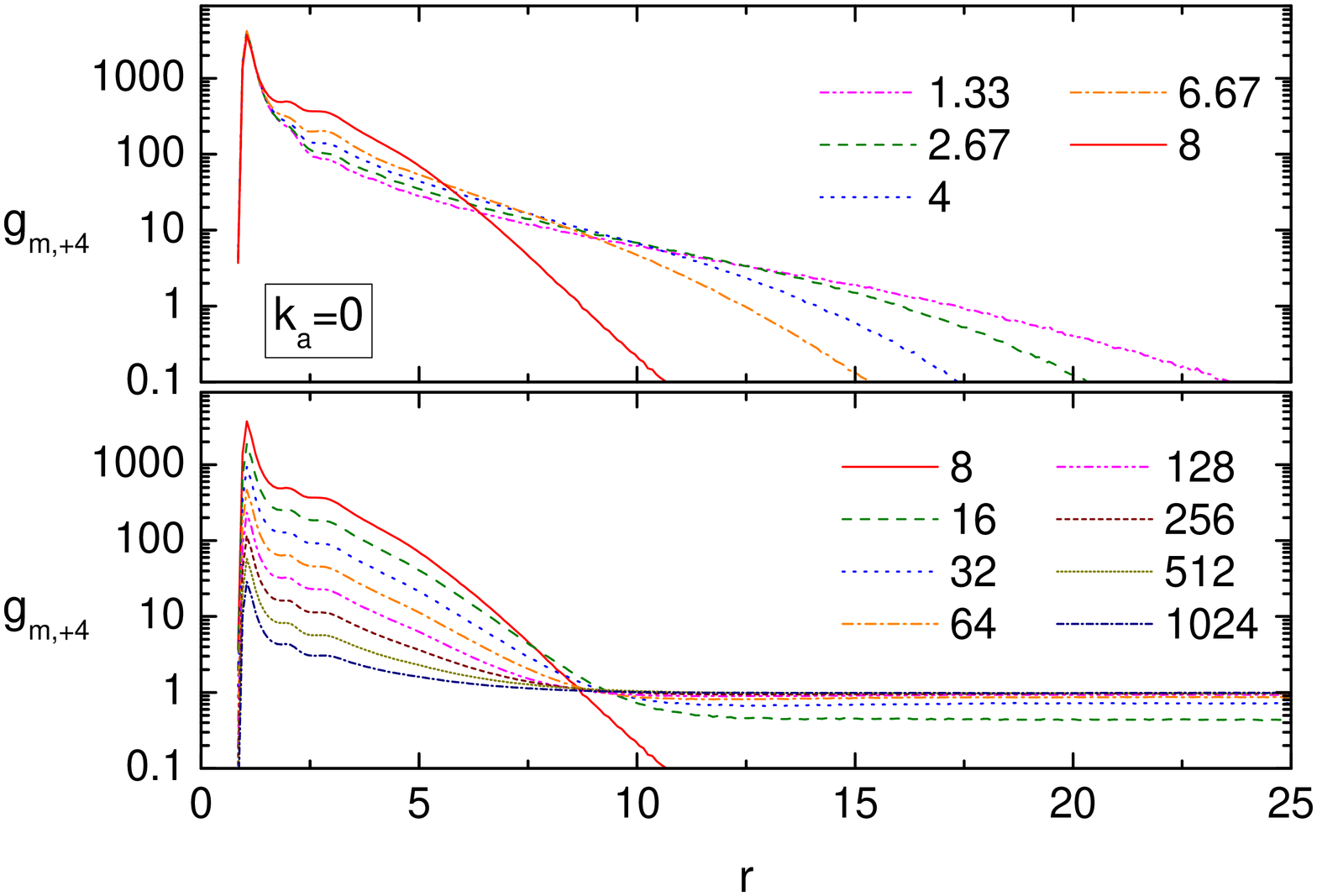}
\caption{$g_{m,+4}(r)$ for (a) the rigid chain, (b) the semiflexible chain, and
(c) the flexible chain at different $C_s$. The value of $C_s\times 10^{5}$ is
indicated in the figures.} \label{f:rdf_m4+} \end{figure}

In addition to the ion distributions in the radial and in the axial directions
discussed in the above sections, we study also the radial distribution function
$g_{m,+4}(r)$ between monomers and tetravalent counterions. $g_{m,+4}(r)$ is
defined as the averaged density function $\left<\rho_{m,+4}(r)\right>$ of the
tetravalent counterions around a monomer dividing the mean density of the
tetravalent counterions in the whole system, $C_s$. It is a correlation
function, which provides the information of organization of the tetravalent
counterions surrounding an individual monomer. The results at different $C_s$
are presented in Fig.~\ref{f:rdf_m4+} for the three chain stiffness, $k_a=0$,
$8$, and $100$.

Fig.~\ref{f:rdf_m4+}(a) shows $g_{m,+4}(r)$ for the rigid chain ($k_a=100$).
The distribution functions show a series of peaks located subsequently at
$r=1.0$, $1.8$, $2.9$, $3.9$, $5.0$, .... Since the chain conformation is a
straight line, the results indicate that the condensed tetravalent counterions
sit on the clefts between two adjacent monomers and form triplets with
monomers, as sketched in the cartoon picture (inside the figure), because the peak
positions match well the distances calculated from the picture.  Similar
triplet formation has been reported by other group using monovalent
counterions, instead~\cite{lo08}. When $C_s\le C_s^*$, the distribution functions
overlap with each other. This is because nearly all of the tetravalent
counterions condense on the chain in this salt region.  As a result, the
correlations between monomers and tetravalent counterions are identical.  When
$C_s> C_s^*$, the curves move downward with increasing $C_s$.  The correlation
decreases, owing to the increase of the non-condensed tetravalent counterions
presented in the bulk solutions. 
 
$g_{m,+4}(r)$ for the semiflexible chain ($k_a=8$)  displays subsequent peaks
too, as shown in Fig.~\ref{f:rdf_m4+}(b). However, these peaks are broader and
less pronounced, in comparison with the previous case, because thermal
fluctuations can take effect on chain conformation for this chain stiffness.
Since the chain collapses in the intermediate salt region, the size of the
chain reduces, leading to a faster decay of $g_{m,+4}(r)$ against $r$. 

$g_{m,+4}(r)$ for the flexible chain ($k_a=0$) shows only one peak at $r = 1$
(see Fig.~\ref{f:rdf_m4+}(c)). This peak marks the close contact of the
counterions with a monomer. It is more pronounced than the first peak for the
semiflexible and the rigid chains, suggesting a stronger correlation.
Following the peak, a broad shoulder appears on the right-hand side. The
shoulder narrows down to a finite region when the salt concentration is
increased from $1.33\times 10^{-5}$ to $8\times 10^{-5}$.  This behavior
reflects two facts: (1) the nature of random arrangement of tetravalent
counterions surrounding a monomer when the chain is flexible, (2) the reduction
of chain size in the intermediate salt region.

\section{Conclusions} 
\label{subsec_conclusions}
Using Langevin dynamics simulations, we have studied ion and charge
distributions in PE solutions from various points of view. The discussed topics
focused on the effects of the two parameters: the salt concentration and the
chain stiffness. The salt concentration ran over a wide range of value in which
the reentrant condensation takes place, and the chain stiffness covers the
three types of behavior discussed in Section~\ref{subsec_chain_size}.  The
results showed that the ion and the charge distributions depend strongly on the
morphology of a chain, while the morphology is determined by both of the two
parameters. When the salt concentration is low and cannot induce collapse of
chain, the ion distribution shows similar profile for flexible, semiflexible
and rigid chains, because these chains all display extended structures. In the
middle salt region, the chain charge is nearly neutralized by the condensed
multivalent counterions. The chains show drastic conformational change. The
flexible chains are collapsed into disordered globule structures whereas the
semiflexible chains are collapsed into ordered structures such as toroid and
hairpin. For the rigid chains, the chain conformation remains rodlike. Since
the chain morphologies are different for these three kinds of chain stiffness,
the ion distributions show significant difference. A general rule is that the
larger the space that a chain structure occupies, the more the ions can
condense on a chain. Therefore, the number of the condensed ions and the
effective chain charge for the semiflexible chains are upper- and lower-bounded
by those curves for the flexible and the rigid chains, respectively.  In the
high-salt region, the condensed multivalent counterions overcompensate the
chain charge on the chain surface.  The flexible and the semiflexible chains
reexpand in the solutions. The latter chain displays a rodlike structure, more
extended than the former one which displays a coil structure. Therefore, the
semiflexible and the rigid chains show similar ion distributions,
distinguishable to the flexible chain. The study of the condensed ion
distribution in the chain axial direction tells us other information. At low or
at high salt concentrations, the condensed ions distribute uniformly on the
chain backbone (because the chain is not collapsed), whereas at intermediate
salt concentrations, the distribution showed specific profile which has direct
connection with the structure of the collapsed chains.  Finally, in the study
of the radial distribution function, we demonstrated that a condensed
multivalent counterion favors to form a triplet with two adjacent monomers.
Consequently, the radial distribution of multivalent counterions shows a series
of peaks away from a monomer. 
   
Before the closure of this paper, we describe more about how flexible chains
and semiflexible chains get reexpanded in a high-salt region.  According to
Fig.~\ref{f:tube_rc3}, the number of the condensed multivalent counterions
increases monotonically when the salt concentration is increased. We know that
flexible chains and semiflexible chains are both collapsed and
charge-neutralized at the equivalence point $C_s^*$.  Further condensing the
multivalent counterions leads to the arising of the repulsive forces, including
the Coulomb repulsion and the excluded volume interaction, in the interior of
the collapsed chain structures. The collapsed chains are therefore getting
looser and becomes more and more unstable.  For a flexible chain, the chain
unfolds gradually into a coil structure. Similar results have been reported
recently in the simulation study of other flexible charged systems, such as
dendrimers~\cite{tian09} and spherical polyelectrolyte brushes~\cite{ni08}.  In
these studies, the dendrimers and the PE brushes were collapsed by multivalent
salt and oppositely charged linear PEs, respectively.  When the concentration
of these condensing agents is higher than the equivalence point, the collapsed
structures reswell due to the excluded volume and the repulsive interactions of
the condensing agents excessively adsorbed on the systems. The reswelling goes
in a continuous way~\cite{wei07}. For a semiflexible chain, there exists, in
addition, a restoring force, owing to the intrinsic chain stiffness, which
plays a crucial role in the chain reexpansion. The binding of chain segments in
a collapsed chain structure decreases when the excess number of the condensed
counterions increases.  Once the restoring force can overcome the binding
force, the chain reexpands in a sudden and drastic way. A first-order
transition takes place~\cite{wei07}. The detailed information
provided in this study gives valuable insight to the problem of DNA
condensation and can help researchers to develop new models to explain these
phenomena more quantitatively.

\section{Acknowledgments}
This material is based upon work supported by the National Science Council, the
Republic of China, under Contract No.~NSC 95-2112-M-007-025-MY2 and NSC
97-2112-M-007-007-MY3.  Computing resources are supported by the National
Center for High-performance Computing under the project ``Taiwan Knowledge
Innovation National Grid''. 


\end{document}